\documentclass[oneside,english]{elsart}
\pdfoutput=1
\usepackage[T1]{fontenc}
\usepackage[latin1]{inputenc}
\usepackage{graphicx}
\usepackage[authoryear]{natbib}

\makeatletter

\providecommand{\tabularnewline}{\\}

\makeatother

\usepackage{babel}

\begin{document}
\begin{frontmatter}

\title{Extinction in a self-regulating population with demographic and environmental
noise}

\author{Alexei J. Drummond$^{(1)}$, Peter D. Drummond$^{(2)}$}

\end{frontmatter}
(1)The University of Auckland, Auckland, New Zealand; \\
(2)The University of Queensland, Brisbane, Queensland, Australia.

\begin{abstract}
We present an explicit unified stochastic model of fluctuations in
population size due to random birth, death, density-dependent competition
and environmental fluctuations. Stochastic dynamics provide insight
into small populations, including processes such as extinction, that
cannot be correctly treated by deterministic methods. We present exact
analytical and simulation-based results for extinction times of our
stochastic model and compare the different effects of environmental
stochasticity and intrinsic demographic stochasticity. We use both
the discrete master equation approach and an exact mapping to a Fokker-Planck
equation (the Poisson method) and stochastic equation, showing they
are precisely equivalent. We also calculate approximate extinction
times using a steepest descent method. This model can readily be extended
to accommodate metapopulation structure and genetic variation in the
population and thus represents a step towards a microscopically explicit
synthesis of population dynamics and population genetics.
\end{abstract}
Keywords: discrete logistic model; logistic growth; self-regulating
population; stochastic population dynamics; stochastic discrete logistic
model; time to extinction

\section{Introduction}

Models of population dynamics have been the subject of much study
in biology. The essay of \citet{Malthus:1798} first suggested exponential
growth. The idea of quantitative limits on growth is usually attributed
to \citet{Verhulst:1838}, who pioneered the use of the \emph{logistic
equation} (LE), and to \citet{Lotka:1920} and \citet{Volterra:1926},
who considered models of predators and prey. Typically these early
studies of population dynamics regards the units of interest as the
number of individuals in populations of one or more species, without
regard to genetic diversity within a population. 

Given an initial focus on large population sizes, this early work
focused on deterministic equations with continuous rather than discrete
variables. A related approach is the \emph{logistic map} (LM) which
is also deterministic and continuous, but treats time as a series
of discrete generations \citep{May:1976qv}. Continuous population
equations like this can be extended to include external \emph{environmental}
fluctuations. For continuous time, this results in a continuous \emph{stochastic
logistic equation} (SLE) used to model populations: typically for
ecological studies. While it is possible to extend continuous state
models to incorporate genetics, they are limited by the fact that
reproduction and mutation have an intrinsically random nature, requiring
a statistical treatment of a discrete jump process.

The \emph{discrete logistic equation} (DLE) was introduced by \citet{Feller:1939}.
This is a continuous time Markov process with discrete states. It
describes the probabilistic time-evolution of a discrete rather than
continuous population, in which the rates of birth or death are linear
or quadratic in the total population number. This model and variations
of it have been the subject of study by a number of researchers, mostly
in the context of calculating extinction times \citep{Nasell:2001uf,Matis:2003,Matis:2004fy,Newman:2004hc,Doering:2005}.
One of the principle virtues of the model is that it can treat \textit{demographic
noise}: the inevitable stochasticity intrinsic to discrete random
events such as birth and death. 

A particular special case of stochastic logistic growth is obtained
when density dependence occurs only through asymptotically decreasing
birth rates, such that a hard upper limit in population size (known
as the maximum carrying capacity) can never be exceeded \citep{Dushoff:2000it}.
This model originally arose in mathematical epidemiology as the closed
stochastic SIS model \citep{Weiss:1971,Nasell:1996bd}; a simple model
of endemic disease in which the maximum carrying capacity is the (constant)
size of the host population. It is also possible to construct a discrete
logistic equation that has density dependent death rates, which is
the approach we use here.

Any continuous-time discrete-state Markov process of this type can
be canonically described by a \textit{birth-death master equation}
and such equations are widely used in chemistry and physics \citep{Gardiner:2004bk}.
An approximate way to treat Master equations - introduced by Kramers
and Moyal - is to replace the discrete variables (like population
size) by continuous variables undergoing a diffusion process\citep{Kramers:1940,Moyal:1949}.
With truncation, this results in a Fokker-Planck equation. After transforming
to a continuous variable stochastic equation, this is simply equivalent
to the SLE \citep{Feller:1951}. 

This approximation results in a form of the SLE which can describe
the effects of demographic noise in the limit of large population
numbers. Thus, the Kramers-Moyal approximation to the discrete logistic
equation is similar to the stochastic logistic equation for environmental
noise, but with a different interpretation of the noise source. It
has the drawback that it is not accurate for small populations. This
leads to exponentially large errors\citep{Gaveau:1996} for important
problems like extinction times, which necessarily involve a small
population. Nevertheless, many published works in mathematical ecology
and epidemiology that address demographic stochasticity have used
the Kramers-Moyal or even more serious types of approximation.

At the same time, quantitative statistical models of the \emph{genetic}
evolution of populations are based on the pioneering work of population
geneticists Fisher\citep{Fisher:1918,Fisher:1930}, \citet{Wright:1930gl}
and \citet{Haldane:1932wy}, whose combined work has laid the foundation
for subsequent research in this area. \textit{\emph{However, many
important questions are inaccessible by current techniques, because
most population genetic theory is based on the idealized Wright-Fisher
model of population and the coalescent \citep{Kingman:1982}, neither
of which handle small randomly fluctuating populations. Thus, the
fields of mathematical ecology and epidemiology, where questions about
extinction, population viability \citep{Boyce:1992,Lande:1993bv,Foley:1994qr,Engen:2000mw,Hakoyama:2000uf,Drake:2006ss,Cairns:2007fu}
and short-term population dynamics are important, have developed quite
independently from theoretical population genetics.}} 

Recently there has been renewed interest in unifying these fields
from two different directions. Firstly researchers interested in the
study of viral evolutionary dynamics have long realized that epidemiological
dynamics and population genetics have overlapping time scales in viruses
\citep{Pybus:2001cl,Pybus:2003hq,Kelly:2003xw,Grenfell:2004jl}. Secondly
and more recently there have been a number of attempts to develop
general theoretical results aimed at extending classical population
genetics concepts and results to the analysis of self-regulating stochastic
population models. Notable examples include calculation of the effective
population size of fluctuating populations that have both environmental
and demographic stochasticity \citep{Engen:2007rq}; development of
a branching-process analogue of the stochastic logistic model with
linear birth rates and density-dependent death rates \citep{Lambert:2005bq}
and its application to the analysis of fixation probabilities in statistically
varying populations \citep{Lambert:2006jo,Champagnat:2007mw}; and
analysis of the one-locus two-allele fixation probability in closed
stochastic SIS model \citep{Parsons:2007ay}.

Similar underlying mathematical issues to those faced by biologists
have been known in the physical sciences for some time. Efficient
techniques for solving master equations directly by simulation were
introduced by \citet{Gillespie:1977}. An \textit{exact} transformation
of master equations to a continuous Fokker-Planck equation and hence
a stochastic form was introduced by \citet{Gardiner:2004bk}, and
is known as the Poisson representation. This approach differs from
the commonly used Kramers-Moyal approximation in that it is exact
for small numbers, provided boundary terms vanish, and hence can be
used to reliably calculate extinction times (among other things).
The more recently developed stochastic-gauge \textit{\emph{Poisson
representation\citep{Deuar:2002,DrummondDeuar:2003,DrummondPD:2004}}}
eliminates boundary term errors that can occur in the Poisson method,
thus making the technique more generally useful.

\textit{\emph{In this paper, we apply these theoretical techniques
to the problem of calculating and simulating extinction times in the
discrete logistic equation with environmental noise added.}} We consider
a logistic model that includes demographic noise due to birth, death,
density-dependent intraspecific competition and a simple model of
environmental noise. We obtain exact analytic results in this equation
which unifies the DLE and SLE models. For pure demographic noise,
we have verified that our exact results agree precisely with a Monte-Carlo
simulation of the master equation, showing that the discrete jump
and continuous noise approaches are exactly equivalent with these
techniques. We also compare our results with direct simulations of
stochastic equations and with steepest-descent approximations. These
provide useful analytic expressions, valid in the limit of large populations.

We present asymptotic results demonstrating that while large carrying
capacity increases extinction times exponentially, it is not the only
factor involved. There is also an exponential dependence on the reproductive
ratio. For a given growth rate and carrying capacity, populations
are exponentially shorter-lived at small reproductive ratio, as $R\rightarrow1$
.

In a following paper, we will extend these results to show that spatial
connectedness provides a robustness against global extinction of a
population. We do this by describing the relationship between the
rate of extinction and the size and connectedness of the metapopulation.
This is analysed quantitatively using a combination of analytic and
numerical techniques. Extensions to include genetics will be treated
elsewhere.

\section{Methods}

In this paper we will revisit and extend the logistic model of population
dynamics. In its original form this model simply assumed that while
small populations may grow exponentially, this growth saturates due
to nonlinear effects such as density-dependent competition for resources.
Thus, for a time-dependent population $N\left(t\right)$ of a given
species,

\begin{equation}
\frac{\partial N}{\partial t}=N\left(g-cN\right),\end{equation}
 where $g$ is the initial growth rate, $c$ describes competition,
and $N_{c}=g/c$ is the deterministic carrying capacity. This is known
as the logistic equation \citep{Verhulst:1838}. If the initial population
values are small, it has the simple property of an initial growth
followed by a gradual approach to the steady-state, in which $N\rightarrow N_{c}$.
Without the competition term there is an unrealistic exponential growth
that fails to account for the inevitable competition for limited resources
that occurs in all natural populations.

\subsection{Environmental and demographic stochasticity}

Nothing is more certain than uncertainty. All populations experience
random fluctuations in numbers. These fluctuations are intrinsic to
the random timing of population changing events, together with the
granularity of populations that are made up of a finite number of
individuals. To describe this reality, it is necessary to use master
equations that deal with discrete outcomes, rather than continuous
rate equations. We wish to analyse this \emph{demographic} stochasticity,
as a measure of minimum, irreducible randomness which cannot be eliminated
from real population dynamics. In nature, there is also randomness
in the birth and death rates, caused by the chaotic fluctuations in
any real environment. Obviously, this \emph{environmental }stochasticity
represents a large part of observed stochasticity, and should not
be ignored. 

Traditional deterministic population equations may include environmental
noise. However, they frequently ignore demographic stochasticity,
even though this is essential to the analysis of extinction rates,
where the discrete difference between zero and one is important. Conversely,
previous treatments of the discrete stochastic logistic model have
generally ignored external fluctuations. An environmental fluctuation
may change the death rate in a population, but at close to zero population
the demographic statistical fluctuations become more important. These
determine the minimum survivable population, before extinction becomes
a near certainty.

The standard treatment of genetic drift also includes stochastic fluctuations
in the frequencies of different genotypes, but ignores such fluctuations
in the overall number of individuals in the population. Our model
can be extended to include both effects and thus provides a platform
for further development in this new area \citep{Parsons:2007ay}.

\subsection{Master Equations}

One can show that there is a stochastic, or random equation that corresponds
exactly - both in terms of averages and fluctuations - to the types
of process we are considering here. Such equations were first considered
in the context of the logistic model, by Feller. Generically, they
are known as birth-death master equations. We first note that all
of the number fluctuations we are interested in can be understood
by combining the genotype and location labels together into one abstract
species $X_{z}$. 

Here, the combined label $z=\{j,\mathbf{x\}}$ indicates genotype
$j$ at location $\mathbf{x}$. Thus, there are a set of discrete
position labels $\mathbf{x}$, each indicating a cell or lattice site
within which there is relatively strong mixing. For convenience, we
label these abstract combinations of location and species in this
section simply as $X_{1}$, $X_{2}$ , and so on.

To derive the stochastic equations, first consider the most general
kinetic equation we are interested in. As a simple model for stochastic
birth/death events, we consider a binary reaction \begin{equation}
\varepsilon_{1}X_{1}+\varepsilon_{2}X_{2}\,\,\,\begin{array}{c}
k(t)\\
\rightarrow\end{array}\,\,\,\varepsilon_{3}X_{3}+\varepsilon_{4}X_{4}\,\,,\end{equation}
which describes $\varepsilon_{1}$ individuals of type $X_{1}$ combining
with $\varepsilon_{2}$ individuals of type $X_{2}$, to give $\varepsilon_{3}$
individuals of type $X_{3}$ and $\varepsilon_{4}$ individuals of
type $X_{4}$. The rate for the process is $R(t)$, which may be time-dependent
or stochastic. There is a birth-death master equation whose solution
gives the time-evolution of probabilities $P\left(\mathbf{N}\right)$,
for finding the total number of individuals of each type equal to
$\mathbf{N}=\left(N_{1},..,N_{D}\right)$ . Here $D$ is the genetic/spatial
dimension, equal to the number of distinct genotypes multiplied by
the number of lattice sites under consideration. 

This master equation is obtained by summing over all processes, where
each binary reaction generates a term of the form:\begin{eqnarray}
\frac{d}{dt}P\left(\mathbf{N}\right) & = & -k(t)N_{1}^{\varepsilon_{1}}N_{2}^{\varepsilon_{2}}P\left(\mathbf{N}\right)\nonumber \\
 & + & k(t)\left(N_{1}'\right)^{\varepsilon_{1}}\left(N_{2}'\right)^{\varepsilon_{2}}P\left(\mathbf{N}'\right)\end{eqnarray}

Here, $\mathbf{N}'=\left(N_{1}+\varepsilon_{1},N_{2}+\varepsilon_{2},N_{3}-\varepsilon_{3},N_{4}-\varepsilon_{4},..,N_{d}\right)$,
with obvious modifications in cases of identical labels. This is difficult
to solve as it stands, due to the large dimensionality of the space
of distributions over integer populations. This complexity increases
exponentially with the dimensionality $D$, which means that both
the vector $P\left(\mathbf{N}\right)$ and the matrix that describe
its microscopic changes, rapidly become too large to be stored or
numerically analysed in the memory of any computer. 

However, there are several techniques for solving these equations
which make use of mappings into probabilistic Markov processes or
stochastic equations. These do not have exponential complexity, but
rather trade off extremely large matrices in favour of sampling over
a probabilistic distribution. While this introduces sampling errors,
the amount of error can be controlled and reduced by increasing the
number of samples. In some cases, exact or approximate analytic solutions
can be obtained without requiring simulations.

Two methods we will use here are the direct method of Gillespie and
the Poisson representation.

\subsection{Direct Gillespie Simulation}

The direct method of simulating stochastic rate equations was originally
applied to the problem of simulating chemical kinetics by \citet{Gillespie:1977}.
Applied to population dynamics this technique is essentially an in-silico
direct model of the population. In this method a single event is one
reaction process, each event adding or subtracting an individual after
a random time interval whose mean is the inverse of the reaction rate.

The Gillespie algorithm produces a single forward-time realization
of the stochastic process of interest. In order to obtain statistics
on the average time-evolution of the process, many replicates are
simulated, each starting with a different random number seed so that
a sample of realisations is obtained that represent the range of outcomes.
An individual Gillespie simulation proceeds one event at a time. The
events considered in this paper are birth, death, and death by density-dependent
competition, although extensions to more general models are easily
made. 

The Gillespie method is straightforward to implement and provides
a useful base-line for comparisons, especially when demographic noise
is the main source of fluctuations. Its main limitations are that
it cannot be used to model large populations due to its computationally
intensive approach, and the addition of environmental noise is nontrivial.

\subsection{Poisson Representation}

While direct simulations are extremely useful computational tool,
they are not directly amenable to analytic solutions, nor are they
immediately suitable for describing a unified theory that combines
demographic and environmental noise. An efficient alternative technique
for this purpose is to use the Poisson representation\citep{ChatGard77,GardChat77,ChatGard78,Gardiner:2004bk}
- an exact expansion over Poisson distributions - to map these equations
into a stochastic differential equation for a continuous Poisson amplitude. 

This proceeds by expanding $P\left(\mathbf{N}\right)$ in a basis
of elementary multivariate Poisson distributions, so that:

\begin{equation}
P\left(\mathbf{N}\right)=\int d[\vec{{x}}]\, f\left(\vec{{x}}\right)P\left(\mathbf{N};\vec{{x}}\right)\,\,,\end{equation}

where $P\left(\mathbf{N}\left|\vec{{x}}\right.\right)$ is the Poisson
distribution, with:

\begin{equation}
P\left(\mathbf{N};\vec{{x}}\right)=\prod_{i}\frac{e^{-{x}_{i}}\left({x}_{i}\right)^{N_{i}}}{N_{i}!}\,\,.\end{equation}
Each elementary process generates an equivalent Fokker-Planck equation
(where $\partial_{i}=\partial/\partial{x}_{i}$):\begin{eqnarray}
\frac{d}{dt}f(t,{\mathbf{x}}) & = & k(t)\left[\left(1-\partial_{3}\right)^{\varepsilon_{3}}\left(1-\partial_{4}\right)^{\varepsilon_{4}}-\left(1-\partial_{1}\right)^{\varepsilon_{1}}\left(1-\partial_{2}\right)^{\varepsilon_{2}}\right]\nonumber \\
 &  & \times{x}_{1}^{\varepsilon_{1}}{x}_{2}^{\varepsilon_{2}}f(t,{\mathbf{x}})\,\,.\label{eq:BasicFPE}\end{eqnarray}

After summing over all elementary processes, one obtains the total
Fokker-Planck equation:\begin{equation}
\frac{d}{dt}f\left(t,\mathbf{x}\right)=\sum_{i}\partial_{i}\left[-A_{i}\left(t,\mathbf{x}\right)+\frac{1}{2}\sum_{j}\partial_{j}D_{ij}\left(t,\mathbf{x}\right)\right]f\left(t,\mathbf{x}\right)\,.\label{eq:genericFPE}\end{equation}
The utility of the Poisson representation is that it can be used to
directly calculate the factorial moments and correlations, through
the equivalences:\begin{equation}
\langle{x}^{n}\rangle=\langle N(N-1)..(N-n+1)\rangle\,\,.\label{eq:Moments}\end{equation}

\subsection{Stochastic realizations}

The Fokker-Planck equation can also be readily converted into an Ito
stochastic differential equation for direct numerical simulation,
where:\begin{equation}
dx_{i}^{I}=A_{i}\left(t,\mathbf{x}\right)dt+\sum_{j}B_{ij}\left(t,\mathbf{x}\right)dW_{j}\left(t\right)\label{eq:genericItoSDE}\end{equation}
Here, $D_{ij}=\sum_{k}B_{ik}B_{jk}$ , and the terms $dW_{i}$ are
independent real Gaussian noise terms with \begin{equation}
\langle dW_{i}dW_{j}\rangle=\delta_{ij}dt\,\,.\end{equation}
 It is necessary that $D_{ij}$ is positive-definite. This may require
introducing a complex Poisson variable $x_{i}$, together with additional
stochastic gauge terms to ensure stability. Relevant technical details
are presented elsewhere\citep{DrummondPD:2004}. The Ito form of stochastic
equation corresponds most directly to a standard Fokker-Planck equation
form, which is generated directly from the Poisson representation
of demographic noise. This describes a forward step in which the multiplicative
noise term is evaluated at the current time.

There is another commonly used form of stochastic equation. The Stratonovich
equation describes a time-symmetric step, which is the broad-band
limit of a finite band-width physical noise. This is the correct form
in which to introduce environmental noise sources originating from
time-dependent rate fluctuations $R(t)$. The two forms can be converted
from one to the other, since the equivalent Stratonovich equation
has the form:

\begin{equation}
dx_{i}^{S}=\tilde{A}_{i}\left(t,\mathbf{x}\right)dt+\sum_{j}B_{ij}\left(t,\mathbf{x}\right)dW_{j}\left(t\right)\label{eq:genericStratonovichSDE}\end{equation}
where \begin{equation}
\tilde{A}_{i}\left(t,\mathbf{x}\right)=A_{i}\left(t,\mathbf{x}\right)-\frac{1}{2}\sum_{j}B_{kj}\left(t,\mathbf{x}\right)\partial_{k}B_{ij}\left(t,\mathbf{x}\right)\,\,.\label{eq:ItoStratTransformation}\end{equation}
These become formally identical if the noise term is independent of
the phase-space variable $\mathbf{x}$. This case is termed additive
noise or constant diffusion. We will make use of a variable change
to achieve constant diffusion behaviour in the analytic calculations.

\subsection{Population processes on individuals}

We can categorize biologically relevant kinetic processes into the
following list, in which we give the simplest stochastic equation
in each case.

\subsubsection{Transformation}

A transformation is a unary reaction in which one species changes
to another at a constant rate, either through mutation or spatial
motion (recall that the subscripts describe both physical and genetic
space). In practise, this may be catalysed by other organisms or molecules,
or involve additional precursor molecules. As long as these additional
species have a constant concentration, they can be neglected, leading
to the reaction:

\begin{equation}
X_{1}\,\,\,\begin{array}{c}
a\\
\rightarrow\end{array}\,\,\, X_{2}\,\end{equation}

\begin{itemize}
\item Fokker-Planck equation:\begin{equation}
\frac{d}{dt}f(t,{x})=a\left[\partial_{1}-\partial_{2}\right]{x}_{1}f(t,{x})\,\,.\end{equation}

\item Stochastic equation (noise-free):\begin{eqnarray}
d{x}_{1} & = & -a{x}_{1}dt\nonumber \\
d{x}_{2} & = & a{x}_{1}dt\,\,.\end{eqnarray}

\end{itemize}
As an example, the death of an organism will be treated in a simplified
way, where the final product is neglected:

\begin{equation}
X\,\,\,\begin{array}{c}
a\\
\rightarrow\end{array}\,\,\,0\,\end{equation}

\begin{itemize}
\item Fokker-Planck equation:\begin{equation}
\frac{d}{dt}f(t,{x})=a\partial_{x}{x}f(t,{x})\,\,.\end{equation}

\item Stochastic equation (noise-free):
\end{itemize}
\begin{equation}
dx=-a{x}\, dt\,.\end{equation}

None of these processes involve stochastic terms, unless the rates
themselves fluctuate. This implies that, at constant rates, transformation
processes map one Poisson distribution into another Poisson distribution.

\subsubsection{Birth}

Consider birth: a process in which a pre-existing organism $X_{1}$
generates two further organisms $X_{2}$ and $X_{3}$. A special case
of this occurs when all of the species involved are identical. 

\begin{equation}
X\,\,\,\begin{array}{c}
b\\
\rightarrow\end{array}\,\,\,2X\,\end{equation}

\begin{itemize}
\item Fokker-Planck equation:\begin{equation}
\frac{d}{dt}f(t,{x})=b\left[-\partial_{x}+\partial_{x}^{2}\right]{x}f(t,{x})\,\,.\end{equation}

\item Ito stochastic equation:\begin{equation}
dx=b{x}dt+\sqrt{2b{x}}dW\left(t\right)\end{equation}

\end{itemize}
The fact that there is noise in these equations simply implies that
birth is intrinsically noisy, in the sense that it increases local
fluctuations above the Poisson level.

\subsubsection{Competition}

Next, consider competition: a process in which two pre-existing organisms
$X_{1}$ and $X_{2}$ compete (usually for resources), reducing the
total number to just one, $X_{3}$. As in birth, a special cases of
this occurs when all of the species involved are identical: 

\begin{equation}
2X\,\,\,\begin{array}{c}
c\\
\rightarrow\end{array}\,\, X\,\end{equation}

\begin{itemize}
\item Fokker-Planck equation:\begin{equation}
\frac{d}{dt}f\left(t,\mathbf{x}\right)=c\left[\partial_{x}-\partial_{x}^{2}\right]{x^{2}}f\left(t,\mathbf{x}\right)\,\,.\end{equation}

\item Ito stochastic equation:\begin{equation}
dx=-c{x^{2}}dt+ix\sqrt{2c}dW(t)\end{equation}

\end{itemize}
Competition tends to reduce fluctuations \emph{below }the Poisson
level. This results in complex Poisson amplitudes in general, unless
the fluctuations of another process keep the overall variance real.
This equation in its present form is unstable, and would require additional
stochastic gauges to be treated correctly \citep{DrummondPD:2004}.
This issue does not arise in the stochastic logistic model, due to
the additional positive diffusion from the birth term.

\section{Discrete logistic model}

Consider the discrete logistic model of a birth process $X\rightarrow2X$,
together with competition $2X\rightarrow X$, and a death process
$X\rightarrow0$. This type of random process can be represented kinetically
as:

\begin{eqnarray}
X & \begin{array}{c}
a\\
\rightarrow\end{array} & 0\,\,.\nonumber \\
X & \begin{array}{c}
b\\
\rightarrow\end{array} & 2X\,\,,\label{eq:kinetic}\\
2X & \begin{array}{c}
c\\
\rightarrow\end{array} & X\,\,.\nonumber \end{eqnarray}

For small populations the resulting net growth rate is $g=b-a$, which
corresponds to the Malthusian fitness of type $X$. However, we have
a model of fitness which is also density dependent and stochastic
\citep{Feller:1939}. In this model the density-dependence come from
the death by competition. It is also possible to have density-dependent
birth-rates, which gives an extra parameter at the expense of more
complicated stochastic behaviour.

Apart from the growth rate $g$ which sets the relevant time-scales,
this process is described by two dimensionless parameters; $N_{c}=g/c$
is the carrying capacity, and $R=b/a$ is the birth-death ratio, also
called the \emph{reproductive} ratio.

\subsection{Master Equation}

The master equation corresponding to the dynamical behaviour in Eq
(\ref{eq:kinetic}) is:\begin{eqnarray}
\frac{d}{dt}P\left(N\right) & = & -\left(a+b+cN\right)NP\left(N\right)\nonumber \\
 & + & b\left(N-1\right)P\left(N-1\right)\,\,.\nonumber \\
 & + & c\left(N+1\right)^{2}P\left(N+1\right)\nonumber \\
 & + & a\left(N+1\right)P\left(N+1\right)\label{eq:logisticmaster}\end{eqnarray}

This is a set of coupled differential equations which can be treated
directly in a single-species case, provided the total population is
not too large. However, such techniques are difficult to extend to
multiple species problems, due to exponentially increased complexity.
Therefore, in this paper we focus on techniques which can be readily
scaled to treat more complex multi-species cases. 

As an example, we show a graph in Figure 1 of a single Gillespie-type
realization of this jump process, leading to an extinction, together
with the average over many realizations. Note that in this picture
the population $N$ is always discrete, with integer jumps at random
times. 

\begin{figure}
\includegraphics[angle=270,width=1\textwidth]{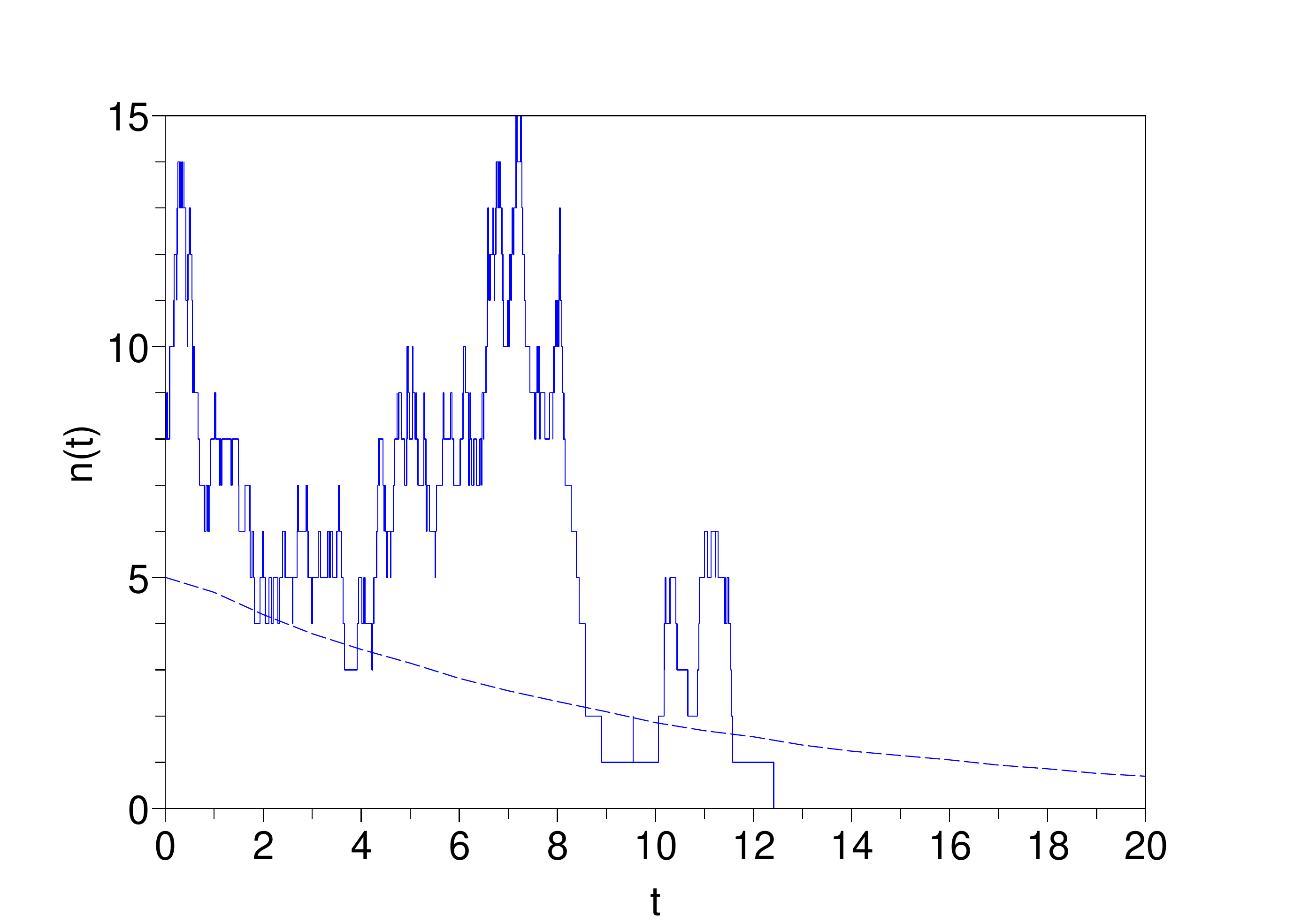}

\caption{Direct simulation results using Gillespie algorithm with parameters
$b=2$, $a=1$, $c=0.2$ . This corresponds to $N_{c}=5$, $R=2$,
$g=1$. The initial condition of $N_{0}=8$ was sampled from a Poissonian
with mean of $x_{0}=5$. The solid line is a single stochastic realization,
showing integer jump behaviour. The dotted line is an average of $10000$
realizations, showing the exponential decline in average population
leading to extinction.}

\end{figure}

\subsection{Poisson Equation}

The master equation (\ref{eq:logisticmaster}) corresponds \emph{exactly}
to the following Fokker-Planck equation in the Poisson representation
for $f\left({x}\right)$ :\begin{eqnarray}
\frac{d}{dt}f\left({x}\right) & = & \frac{\partial}{\partial{x}}\left[-g{x}+cx^{2}+\frac{\partial}{\partial{x}}\left(bx-cx^{2}\right)\right]f\left({x}\right)\,\,.\end{eqnarray}
In terms of the abstract notation of Eq (\ref{eq:genericItoSDE}),
\begin{eqnarray}
A(x) & = & gx-cx^{2}\nonumber \\
D(x) & = & 2(bx-cx^{2})\end{eqnarray}
which implies there is a corresponding Ito stochastic differential
equation:

\begin{equation}
dx^{I}=\left(g{x}-cx^{2}\right)dt+\sqrt{2\left(bx-cx^{2}\right)}dW\,\,,\label{eq:demographicItoEq}\end{equation}

where: \begin{equation}
\langle dW^{2}\rangle=dt\,.\end{equation}
This transforms the discrete master equation given above to a completely
equivalent Ito stochastic differential equation. The equation is valid
in the domain $0\le x\le c/b$ , in which region the noise terms are
all real. As the noise term vanishes at the boundaries, it is appropriate
to use reflecting boundary conditions. The boundary at $x=0$ is absorbing;
once it is reached, the variable $x$ remains at zero. This corresponds
to extinction of the population on this trajectory.

\subsection{Numerical Solutions}

In Figure 2, we show a single stochastic realization of this equation,
together with a time average. Note that here the stochastic variable
$x$ is continuous; it is the mean of one of the Poisson distributions
used in the expansion. Even though there is no direct correspondence
between each realization in Figures 1 and 2, they have identical means.
More general observables can be transformed between these stochastic
methods using the factorial moment equivalences, (\ref{eq:Moments}).

\begin{figure}
\includegraphics[angle=270,width=1\textwidth]{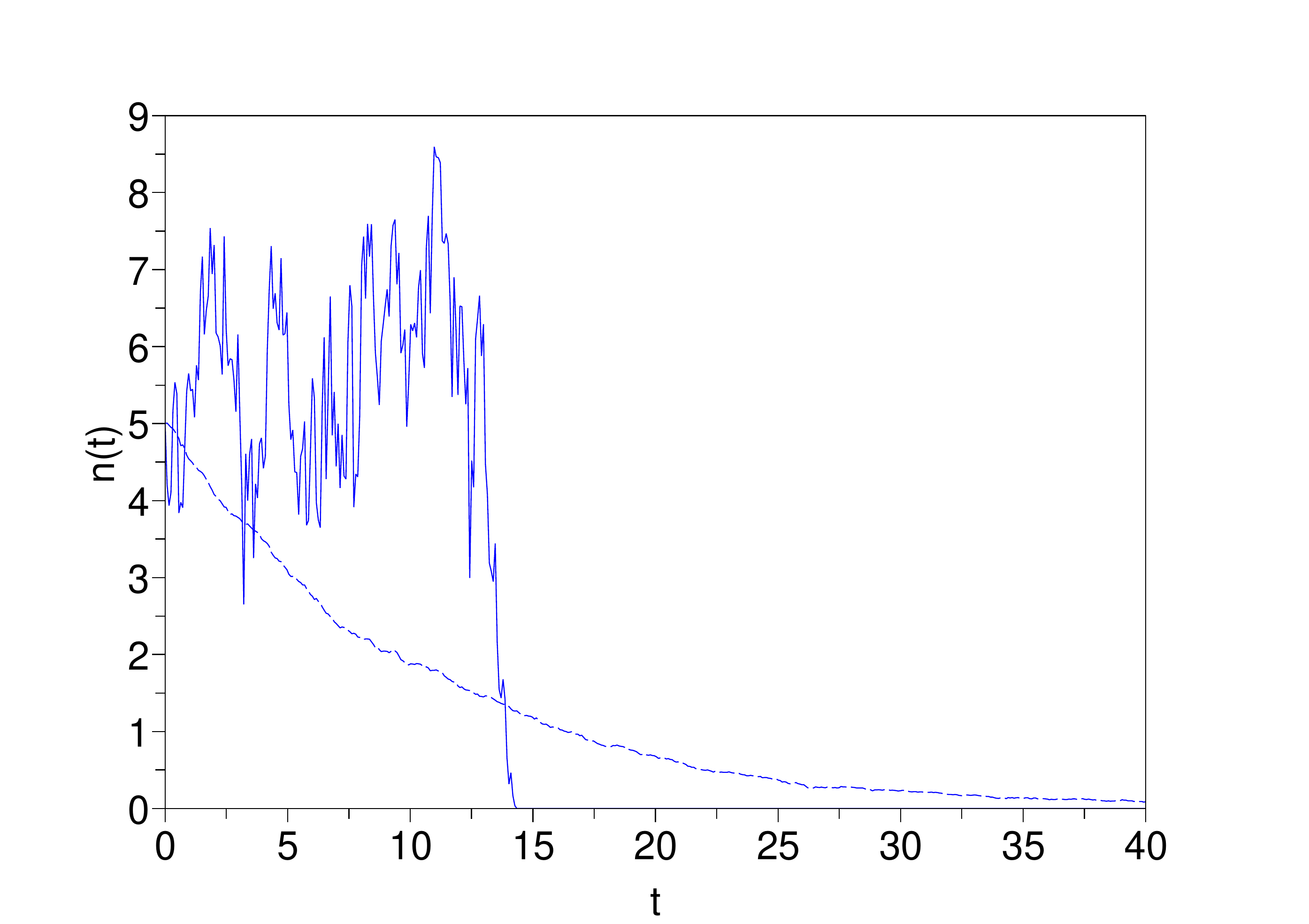}

\caption{Poisson simulation results using Stratonovich equations with parameters
$b=2$, $a=1$, $c=0.2$ or $R=2$, $N_{c}=5$ as in Figure 1. The
initial condition was a Poissonian with mean of $x_{0}=5$. The solid
line is a single stochastic realization, showing continuous stochastic
behaviour. The dotted line is an average of $10000$ realizations,
showing the exponential decline in average population leading to extinction,
just as in the Gillespie result. Integration step-size utilized was
$dt=0.04$, RK4 Runge-Kutta algorithm. Average extinction time $T=10.04\pm0.27$
using Eq(\ref{eq:StochasticExtinctionPoisson}).}

\end{figure}

In these computer simulations, we use an implicit central difference
method with the time-symmetric Stratonovich form of Eq (\ref{eq:genericStratonovichSDE}),
including the transformed drift term of Eq (\ref{eq:ItoStratTransformation}),
again supplemented by absorbing boundary conditions at $x=0$:\begin{equation}
dx^{S}=\left(\left[g+c\right]x-b/2-cx^{2}\right)dt+\sqrt{2\left(bx-cx^{2}\right)}dW\,\,,\label{eq:demographicStratEq}\end{equation}

Points to notice about this approach are:

\begin{itemize}
\item The stochastic equation is for the mean of an 'equivalent' Poisson
distribution
\item $P\left(N\right)$ must be reconstructed from an average over these
Poisson distributions
\item Even if $g=0$ there is noise if $b\neq0$; births cannot cancel deaths
\emph{exactly.} 
\item This intrinsic random noise occurs \emph{in addition to} any environmental
fluctuations 
\item The relative noise size scales as $1/\sqrt{N}$ of the deterministic
terms in the equation 
\item There is no noise if ${x}=0$; this means extinction, as shown by
the Ito form. 
\end{itemize}

\subsection{Population statistics}

In order to demonstrate an important property of the model, consider
the time-evolution of moments, without competition so that $c=0$.
In the stochastic case, moments can be calculated, using standard
Ito variable change \citep{Gardiner:2004bk} rules:\begin{eqnarray}
\langle d{x}^{n}\rangle & = & \langle\left(x+dx\right)^{n}\rangle-\langle x^{n}\rangle\nonumber \\
 & = & n\langle dxx^{n-1}\rangle+\frac{n\left(n-1\right)}{2}\langle dx^{2}x^{n-2}\rangle\nonumber \\
 & = & ng\langle{x}^{n}\rangle dt+n\left(n-1\right)b\langle{x}^{n-1}\rangle\,.\end{eqnarray}

This means that, even if $g=0$, the population distribution does
change statistically. In classical population dynamics, one would
have a stable equilibrium with a constant population. However, demographic
noise modifies this substantially. If we start from a Poissonian with
amplitude ${x}_{0}$, then the factorial moments \emph{increase} with
time even though the mean is constant:\begin{eqnarray}
\langle{x}^{1}\rangle & = & {x}_{0}\nonumber \\
\langle{x}^{2}\rangle & = & {x}_{0}^{2}+2bt{x}_{0}\nonumber \\
\langle{x}^{3}\rangle & = & {x}_{0}^{3}+6\left(bt\right)^{2}{x}_{0}\nonumber \\
\langle{x}^{4}\rangle & = & {x}_{0}^{4}+24\left(bt\right)^{3}{x}_{0}\,\,.\end{eqnarray}

Thus, for example, the standard deviation $\sigma=\sqrt{\langle N^{2}-\bar{N}^{2}\rangle}$
can now be calculated to give:\begin{equation}
\sigma=\sqrt{\bar{N}\left(1+2bt\right)}\,\,,\end{equation}
 This shows that the initial Poissonian variance (equal to the population
mean) increases linearly with time. Some populations become large,
others become extinct -- even if $g$, the growth rate, is zero.

\section{Demographic extinction times}

We wish to analyse the extinction rate of a single isolated population
undergoing discrete logistic population dynamics with competition
included. We will assume an initial Poissonian distribution with mean
$x_{0}$, as this allows us to include some of the effects of initial
population fluctuations in biologically relevant cases. 

There are several equivalent methods to carry out this calculation.
One method is a direct calculation from the master equation, which
gives a result without any environmental noise effects. A direct Gillespie
simulation is also feasible. We will use this as a reference calculation,
since the technique has the advantage that it can be readily extended
to more complex cases. Because the simulations are relatively time-consuming
for large populations and long extinction times, alternative approaches
are desirable.

To obtain analytic results, we will use the Poisson equation. This
has both exact and asymptotic approximate solutions for extinction
times. In the next section we show that the approach can be readily
extended to include environmental noise.

\subsection{Exact extinction times}

Calculating first-passage times for diffusion or Fokker-Planck processes
is a well known problem in physics, and there is an extensive literature.
Here there is an interesting subtlety. Even when the Poisson variable
is non-vanishing so that $x>0$ , there is a fraction of size $e^{-x}$
in the corresponding sample of real populations that is already extinct.
To take account of this, the standard first-passage time calculation
needs modifications. 

We start by defining the rate of extinction at time $t$ given an
initial Gaussian with mean $x_{o}$ as $\mathcal{R}(t|x_{0})$. This
is given by the rate of change of the probability for extinction $\mathcal{E}(t|x_{0})$
given an initial Poisson distribution at $x_{0}$:

\begin{eqnarray}
\mathcal{R}(t|x_{0}) & = & \frac{d}{dt}\mathcal{E}(t|x_{0})\nonumber \\
 & = & \frac{d}{dt}\int\, f\left(t,x|x_{0}\right)e^{-x}dx\,\,.\end{eqnarray}

The average time to extinction $T$ given an initial Poisson distribution
at $x_{0}$ involves a time-integral over the extinction rate:\begin{eqnarray}
T(x_{0}) & = & \int_{0}^{\infty}t\mathcal{R}(t|x_{0})dt\nonumber \\
 & = & \int_{0}^{\infty}t\frac{d}{dt}\mathcal{E}(t|x_{0})dt\nonumber \\
 & = & -\int_{0}^{\infty}t\frac{d}{dt}\mathcal{A}\left(t,|x_{0}\right)dt\,\,.\end{eqnarray}
Here we have introduced an alive probability defined as\begin{eqnarray}
\mathcal{A}\left(t,|x_{0}\right) & = & 1-\mathcal{E}(t|x_{0})\nonumber \\
 & = & \int_{0}^{x_{m}}\, f\left(t,x|x_{0}\right)\left[1-e^{-x}\right]dx\,\,.\end{eqnarray}
The absorbing state at $x=0$ means that the distribution decays to
a delta function at long times: $f\left(\infty,x|x_{0}\right)=\delta(x)$,
so that:

\[
\lim_{t\rightarrow\infty}t\mathcal{A}\left(t,|x_{0}\right)=0\,.\]
This last result allows a further simplification from integration
by parts, so that:\begin{equation}
T(x_{0})=\int_{0}^{\infty}\mathcal{A}\left(t,|x_{0}\right)dt\,\,.\end{equation}

Since $f\left(t,x|x_{0}\right)$ is a normalized, probabilistic propagator
function in the Poisson representation, this leads to a simple prescription
for calculating extinction times in a numerical simulation:\begin{equation}
T(x_{0})=\int_{0}^{\infty}\left\langle 1-e^{-x}\right\rangle dt\,.\label{eq:StochasticExtinctionPoisson}\end{equation}

Analytically, however, we can do better, and even find an exact solution
just requiring numerical integration. $\mathcal{A}$ is a linear functional
of the Poisson distribution $f\left(t,x|x_{0}\right)$, and as such
must satisfy the well known 'backward' Kolmogorov equation \citep{Gardiner:2004bk}
in terms of its initial condition:\begin{equation}
\frac{d}{dt}\mathcal{A}(t|x_{0})=\left[A(x_{0})\partial_{x_{0}}+\frac{1}{2}D\left(x_{0}\right)\partial_{x_{0}}^{2}\right]\mathcal{A}(t|x_{0})\,.\end{equation}

We can now integrate the above equation over all times to obtain an
ordinary differential equation for $T\left(x_{0}\right)$. Initially,
$\mathcal{A}(0|x_{0})=1-e^{-x_{0}}$ so that: \begin{eqnarray}
\mathcal{A}(\infty|x_{0})-\mathcal{A}(0|x_{0}) & = & e^{-x_{0}}-1\nonumber \\
 & = & \left[A(x_{0})\partial_{x_{0}}+\frac{1}{2}D\left(x_{0}\right)\partial_{x_{0}}^{2}\right]T(x_{0})\,.\label{eq:Extinct-time}\end{eqnarray}
We note the following boundary condition on the extinction time: $T\left(0\right)=0$.
This simply means that a population starting with a zero Poisson mean
is extinct immediately, as can also be verified from the absorbing
boundary condition on $\mathcal{A}$. This condition can be utilized
to solve the extinction time differential equation, Eq (\ref{eq:Extinct-time}). 

We first introduce an auxiliary function, \begin{eqnarray}
\psi(x_{0}) & = & \frac{2}{D(x_{0})}\exp\left[\int^{x_{0}}\frac{2A(x)}{D(x)}dx\right]\,,\nonumber \\
 & = & \frac{1}{x_{0}\left(b-cx_{0}\right)}\exp\left[\int^{x_{0}}\left(1-\frac{a}{b-cx}\right)dx\right]\nonumber \\
 & = & \frac{e^{x_{0}}}{x_{0}}\left(b-cx_{0}\right)^{\nu}\,,\end{eqnarray}
where $\nu\equiv a/c-1$. In terms of $\psi,$ Eq (\ref{eq:Extinct-time})
has the particular solution: \begin{eqnarray}
T\left(x_{0}\right) & = & 2\int_{0}^{x_{0}}\frac{dx}{D(x)\psi(x)}\int_{x}^{x_{m}}\psi(z)\left(1-e^{-z}\right)dz\,.\label{eq:ExtinctionTime}\end{eqnarray}

This satisfies the backward Kolmogorov equation, and the boundary
condition at $x=0$. In principle, another solution is possible, as
the differential equation is of second order. However, Eq (\ref{eq:Extinct-time})
has a regular singular point at $x=x_{m}$, whose indicial equation
indicates that any other independent solution of the homogeneous equation
would be singular at $x=x_{m}$, and hence inadmissable. 

This result is remarkably similar to the standard expression for a
first-passage time of a diffusion process \citep{Gardiner:2004bk}
except for the factor of $\left(1-e^{-z}\right)$ in the integrand.
This has the simple intuitive interpretation that it projects out
the fraction of the population in a given Poisson ensemble that is
still `alive', ie, has not yet reached extinction. While this term
is essential if an exact result is required, it is typically a small
correction in an expression dominated by the exponential factors in
$\psi$. 

Extinction time results for typical parameter values are given in
Table (\label{Extinction-time-results}), compared to results for
direct Gillespie simulations. There is agreement within two standard
deviations of the computational sampling error in all cases. This
provides numerical evidence for the equivalence between the discrete
and continuous variable techniques for calculating extinction times.
Numerical integrations of Eq (\ref{eq:ExtinctionTime}) were checked
for accuracy to at least four significant figures, using an adaptive
routine (tolerance of $.5\times10^{-6}$) for the outer integral,
and a fixed step integration with $4\times10^{3}$ steps of the $\sqrt{z}$
variable to ensure good accuracy at small $z$ values.

\begin{table}
\begin{centering}
\caption{Time to extinction in stochastic logistic population dynamics, relative
to $g=1$.\label{tab:Time-to-extinction}}
\begin{tabular}{|c|c|c|c|c||c|}
\hline 
R & $N_{c}$ & Exact $T_{e}$ & Gillespie $gT_{e}$ & Asymptotic $T_{e}$ & \# Gillespie simulations\tabularnewline
\hline
\hline 
1.2 & 5 & 1.820 & 1.821$\pm$ 0.002 & 5.031 & $10^{6}$\tabularnewline
\hline 
1.5 & 5 & 4.587 & 4.583$\pm$0.004 & 6.580 & $10^{6}$\tabularnewline
\hline 
2.0 & 5 & 10.126 & 10.13 $\pm$ 0.01 & 11.18 & $10^{6}$\tabularnewline
\hline 
3.5 & 5 & 32.91 & 32.97 $\pm$ 0.03 & 32.32 & $10^{6}$\tabularnewline
\hline 
6 & 5 & 83.92 & 83.95 $\pm$ 0.08 & 79.99 & $10^{6}$\tabularnewline
\hline 
1.2 & 10 & 3.996 & 3.990$\pm$ 0.004 & 5.534 & $10^{6}$\tabularnewline
\hline 
1.5 & 10 & 12.86 & 12.84 $\pm$ 0.01 & 11.97 & $10^{6}$\tabularnewline
\hline 
2.0 & 10 & 41.22 & 41.27 $\pm$ 0.04 & 36.66 & $10^{6}$\tabularnewline
\hline 
3.5 & 10 & 291.3 & 291.5 $\pm$ 0.3 & 276.9 & $10^{6}$\tabularnewline
\hline 
6 & 10 & 1430 & 1431 $\pm$ 1.4 & 1399 & $10^{6}$\tabularnewline
\hline 
1.2 & 20 & 10.60 & 10.60$\pm$0.01 & 9.472 & $10^{6}$\tabularnewline
\hline 
1.5 & 20 & 66.03 & 66.06 $\pm$0.06 & 56.09 & $10^{6}$\tabularnewline
\hline 
2.0 & 20 & 593.9 & 589.7$\pm$1.9  & 557.6 & $10^{5}$\tabularnewline
\hline 
3.5 & 20 & $2.835\times10^{4}$ & $2.825\times10^{4}\pm89$  & $2.874\times10^{4}$ & $10^{5}$\tabularnewline
\hline 
6 & 20 & $5.884\times10^{5}$ & $5.936\times10^{5}\pm6.0\times10^{3}$ & $6.053\times10^{5}$ & $10^{4}$\tabularnewline
\hline
\end{tabular}
\par\end{centering}
\end{table}

This set of calculations and simulations investigates the mean time
to extinction for a variety of parameter combinations. The first two
sets of population sizes are relatively small to facilitate direct
comparison between the Gillespie method and exact approach. The Gillespie
results were obtained by simulation of $10^{6}$ realizations for
each of the first twelve parameter combinations. Computational limitations
necessitated reduction of the number of Gillespie simulations for
the last three parameter combinations, although the numerical integration
of the exact result is straightforward. Each simulated realization
started with a population size drawn from a Poisson distribution with
mean of $N_{c}=g/c$. This is also the initial condition for the exact
computations. The extinction time results versus carrying capacity
at different $R$ values are also graphed in Figure 3. Nearly identical
results (within sampling error) were found in a number of stochastic
simulations, using the Stratonovich form to give increased numerical
accuracy.

\begin{figure}
\includegraphics[angle=270,width=1\textwidth]{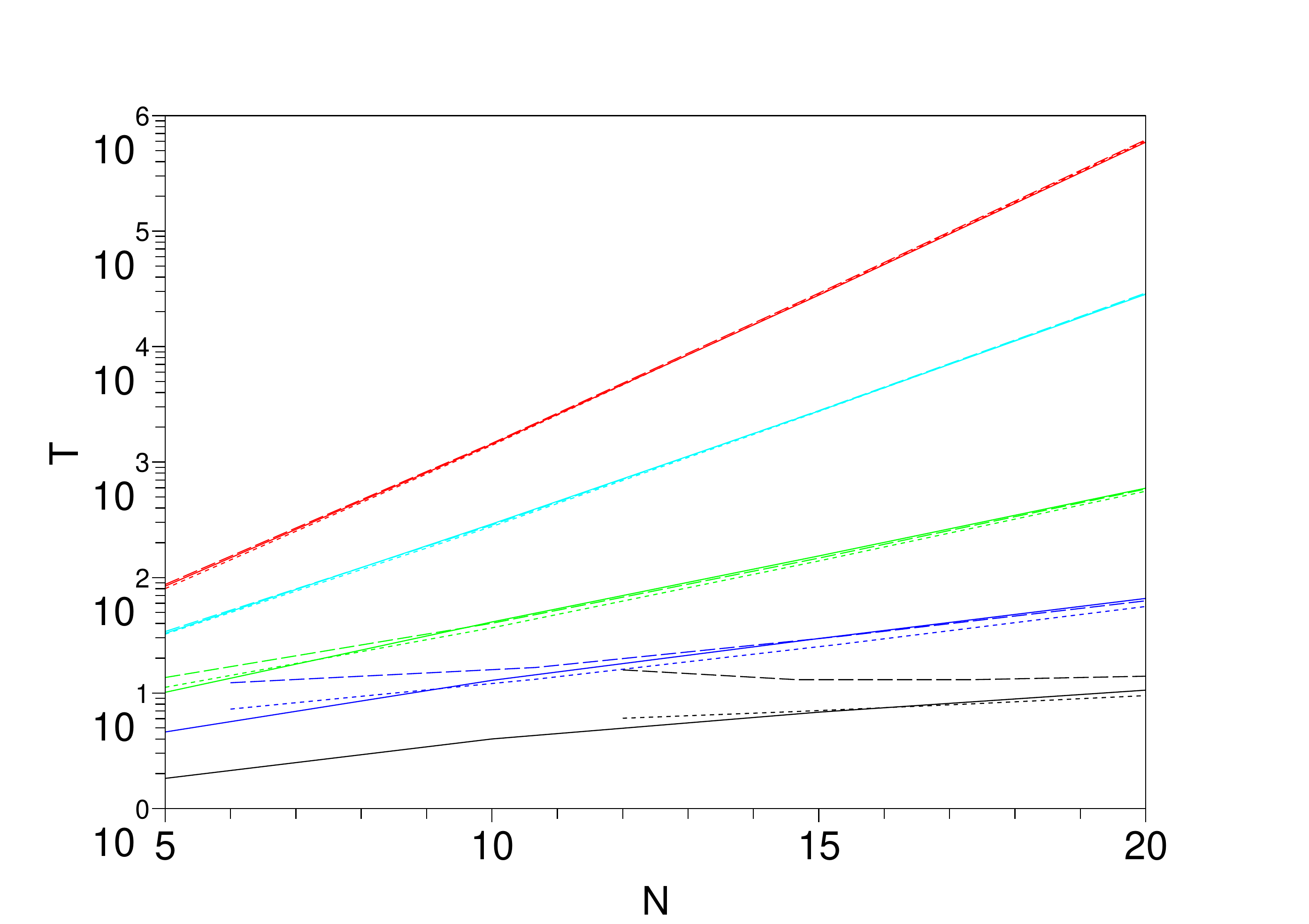}

\caption{Dimensionless time to extinction versus carrying capacity (population)
$N$, for the same range of $R$ values as in the table. Starting
from the lower lines, $R=1.2,1.5,2,3.5,6$. Solid lines are exact
from Eq(\ref{eq:ExtinctionTime}), dashed line are the steepest descent
result from Eq (\ref{eq:DemSteepest}), dotted lines are asymptotic
(large $N$) expressions from Eq (\ref{eq:Asymptotdemographic}).}

\end{figure}

A clear feature of the results is the exponential increase of extinction
times with total population number. In addition, the results show
that there can be changes of just as large a magnitude when the reproductive
ratio $R$ of birth to death rates change. This is graphed directly
in Figure 4.

\begin{figure}
\includegraphics[angle=270,width=1\textwidth]{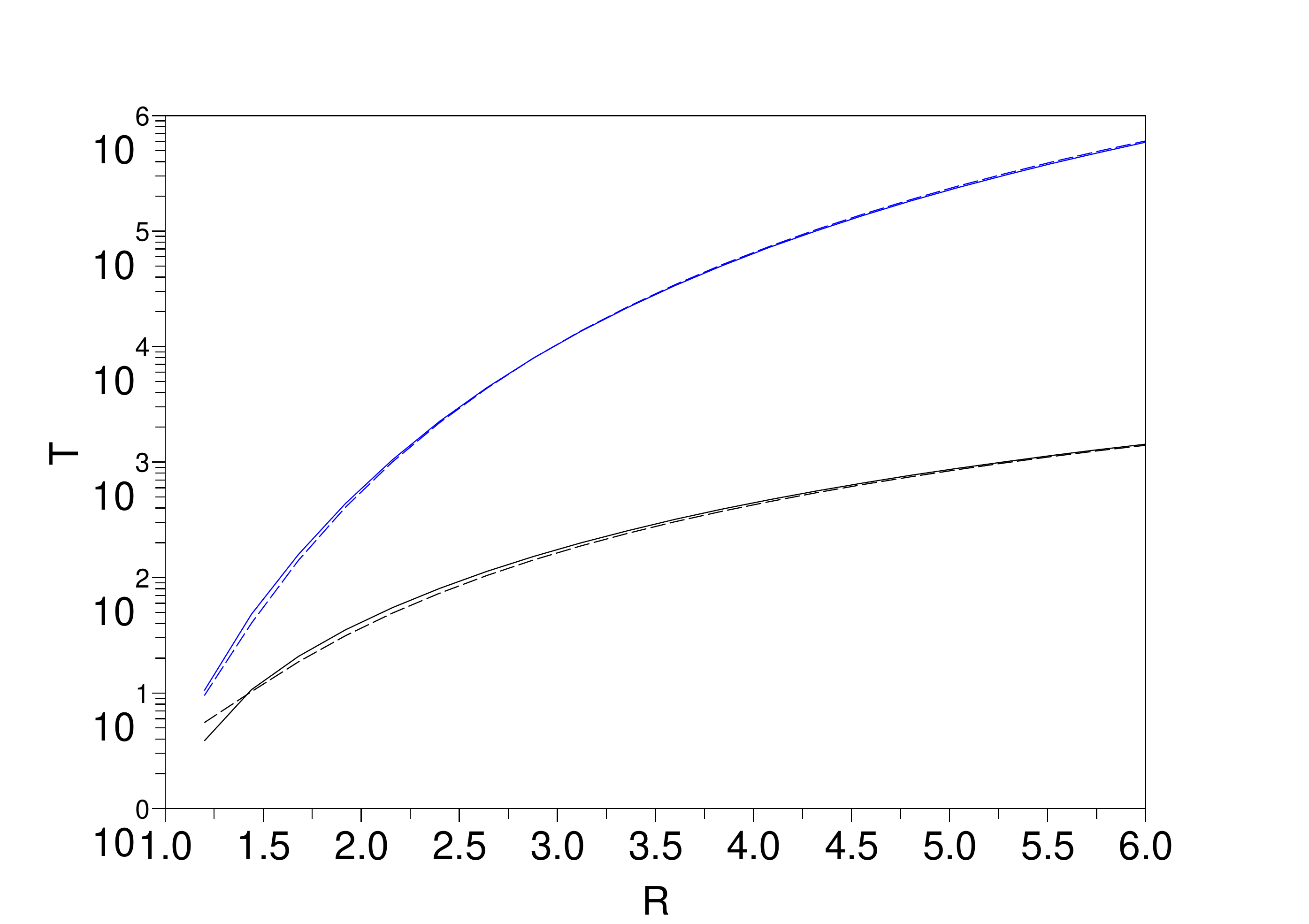}

\caption{Dimensionless time to extinction versus reproductive ratio $R$, for
$N=10,\,20$ (lower and upper curves respectively). Solid line is
the exact result, dashed line is the asymptotic expression (\ref{eq:Asymptotdemographic})}

\end{figure}

These effects will be analyzed in more detail in the remainder of
this section, where we derive the approximate analytic result found
in the last column of the table.

\subsection{Quasi steady-state}

In order to understand these results as a diffusion process, consider
the effect of an artificial reflecting boundary at $x=\varepsilon$.
One can imagine this as being caused by the external intervention
of a benevolent ecologist, wishing to prevent extinction by adding
further individuals when the population becomes too small to be viable.

With this new boundary there is a steady-state equilibrium, which
allows us to derive an effective potential for the Fokker-Planck equation.
The solution is then:\begin{eqnarray}
f_{\infty}\left(x\right) & = & \frac{2\mathcal{N}}{D(x)}\exp\left[\int^{x}\frac{2A(z)}{D(z)}dz\right]\,\nonumber \\
 & = & \mathcal{N}\psi(x)\nonumber \\
 & = & \mathcal{N}\frac{e^{x}}{x}\left(b-cx\right)^{\nu}\,.\end{eqnarray}

Apart from the normalization factor $\mathcal{N}$, the steady-state
distribution is just the auxiliary function used to calculate extinction
times given above. This distribution is not normalizable in the limit
of $\varepsilon\rightarrow0$, which indicates it is not a true steady-state.
With this isolated population model, the steady-state probability
of having $x<\epsilon$ becomes infinite, for arbitrarily small values
of $\varepsilon.$ Thus, while a local equilibrium can be reached
over short times which has a quasi-steady-state behaviour, the only
true steady-state has population zero. 

While we already have a result for the extinction time, we can use
the concept of a quasi-steady-state to develop approximate expressions
which give a better intuitive understanding.

\subsection{Approximate extinction times}

This quasi-steady-state distribution is typically double-peaked. Some
populations are near the deterministic steady-state, while some populations
are zero. This may be thought of as due to a potential barrier. There
is a finite probability that a population which is locally stable
will penetrate the potential barrier and reach the irreversible state
of $x=0$, which means that an extinction has occurred. We can calculate
the extinction time approximately as an escape probability, from the
deterministic or locally stable value through to extinction at $x=0$.
This is equivalent to a steepest descent approximation \citep{Gardiner:2004bk}
to the exact integral expression given in Eq (\ref{eq:ExtinctionTime}).

To simplify this calculation, we first change to a new variable $y$,
with constant diffusion rate. The variable change is defined so that:\begin{equation}
\frac{dx}{dy}=\Delta=\sqrt{x\left(b-cx\right)}\,\,.\end{equation}
 The Fokker-Planck equation for the corresponding probability distribution
$g(y)=\Delta f(x)$ is:\begin{equation}
\frac{\partial}{\partial t}g(y)=\left[\frac{\partial}{\partial y}V'(y)+\frac{\partial}{\partial y}\right]g\left(y\right)\,,\end{equation}
where $V'(y)=dV/dy$ , and $V(y)$ is the potential for the new distribution.
This is obtained from the quasi-steady-state distribution - since
\begin{eqnarray}
V(y) & = & -\ln\left(\Delta f_{\infty}\right)\nonumber \\
 & = & -x+\frac{1}{2}\ln\left[x\left(b-cx\right)^{1-2d/c}\right]\,\,.\end{eqnarray}
The potential derivative, which defines the turning points, is:\begin{equation}
V'(y)=\frac{1}{\Delta}\left[cx^{2}-(g+c)x+\frac{b}{2}\right]\,\,.\end{equation}

The potential has turning points at $V'(y^{\pm})=0$ , that are given
by solving the corresponding quadratic in the Poisson variable $x$,
so that:\begin{equation}
x^{\pm}=\frac{g+c\pm\gamma}{2c}\,\,,\end{equation}
where $\gamma\equiv\sqrt{(g+c)^{2}-2bc}$. We now follow standard
techniques to calculate first-passage times, and neglect the small
Poisson correction of $(1-e^{-x})$ in the integrand. It is also necessary
to calculate the curvature of the potential, which is:

\begin{eqnarray}
V"(y^{\pm}) & = & \Delta\frac{d}{dx}V'(y^{\pm})\nonumber \\
 & = & \pm\gamma\,\,.\end{eqnarray}
The average extinction time is governed by the potential difference
and curvature. Provided $g\gg\sqrt{2bc}$ , it is given approximately
by:\begin{equation}
T=\frac{2\pi}{\gamma}\exp\left[V(x^{-})-V(x^{+})\right]\,\,.\label{eq:DemSteepest}\end{equation}

\subsection{Limiting behavior}

The table and graphs show an exponential increase in extinction times
with carrying capacity $N_{c}$, as well as a further marked dependence
on the ratio of gain to death rate. Here $R=b/a$ is the birth-death
ratio, also called the \emph{reproductive} ratio, which we introduce
to obtain a dimensionless expression. We can make a further simplification
by calculating the limit of the approximate exponential term at large
carrying capacity, or $c\rightarrow0$ . This gives the limiting result
for the turning points that:\begin{eqnarray}
x^{+} & = & N_{c}=\frac{g}{c}+1-\frac{b}{2g}\nonumber \\
x^{-} & = & b/2g\,.\end{eqnarray}

Next, using these asymptotic values, and evaluating the potentials
at the turning points, we find to leading order that:\begin{eqnarray}
T & = & T_{0}\exp\left\{ N_{c}\left[1-\ln\left(R\right)/\left(R-1\right)\right]\right\} +O\left(1/N_{c}\right)\nonumber \\
 & = & T_{0}\exp\left\{ N_{eff}\right\} \,.\label{eq:Asymptotdemographic}\end{eqnarray}

Here we define a fundamental time-scale of $T_{0}$, and an \emph{effective}
carrying capacity $N_{eff}$ which determines the exponent: \begin{eqnarray}
T_{0} & = & \frac{2\pi R}{ge}\sqrt{\frac{1}{2N(R-1)}}\nonumber \\
N_{eff} & = & N_{c}\left[1-\ln\left(R\right)/\left(R-1\right)\right]\,.\end{eqnarray}
This expression is obtained assuming that $R\ll N_{c}$, since we
are interested here in the limit of large carrying capacity. It is
instructive to consider what happens at large and small relative death
rates, which gives a leading order asymptotic results of:\begin{eqnarray}
\lim_{R-1\ll1}\ln T=N_{eff} & \rightarrow & \frac{1}{2}N_{c}\left(R-1\right)+O\left(\ln N_{c}\right)\nonumber \\
\lim_{R\gg1}\ln T=N_{eff} & \rightarrow & N_{c}+O\left(\ln N_{c}\right)\,.\end{eqnarray}

A comparison of the approximate method with exact results is given
in Table (\ref{tab:Time-to-extinction}). The method performs worst
for the first set of population parameters, which exhibit very rapid
extinction rates. This is due to the fact that many populations in
this regime are below the critical threshold immediately, so don't
have a quasi-steady-state. However the approximation behaves very
well for larger population sizes where a genuine quasi-steady-state
exists as demonstrated by the $N_{c}=20$ simulations, where the approximation
is generally within 10\% of the exact results. 

This approximate result is also compared to the exact calculation
in Figure 5, where the relative error is graphed for three different
$R$ values. Agreement is excellent for large populations, provided
that growth is larger than critical, ie, $g\gg g_{c}=\sqrt{2bc}$
, which means that $N_{c}\gg2R/(R-1)$. If the growth rate is less
than this critical value, the potential has no local minimum, and
extinction will occur on time-scales of the order of the inverse growth-rate
for any initial population. We found no advantage in using the full
steepest descent result of Eq (\ref{eq:DemSteepest}) over the simpler
equation (\ref{eq:Asymptotdemographic}), with the asymptotic result
actually giving smaller errors at low $N$ values.

\begin{figure}
\includegraphics[angle=270,width=1\textwidth]{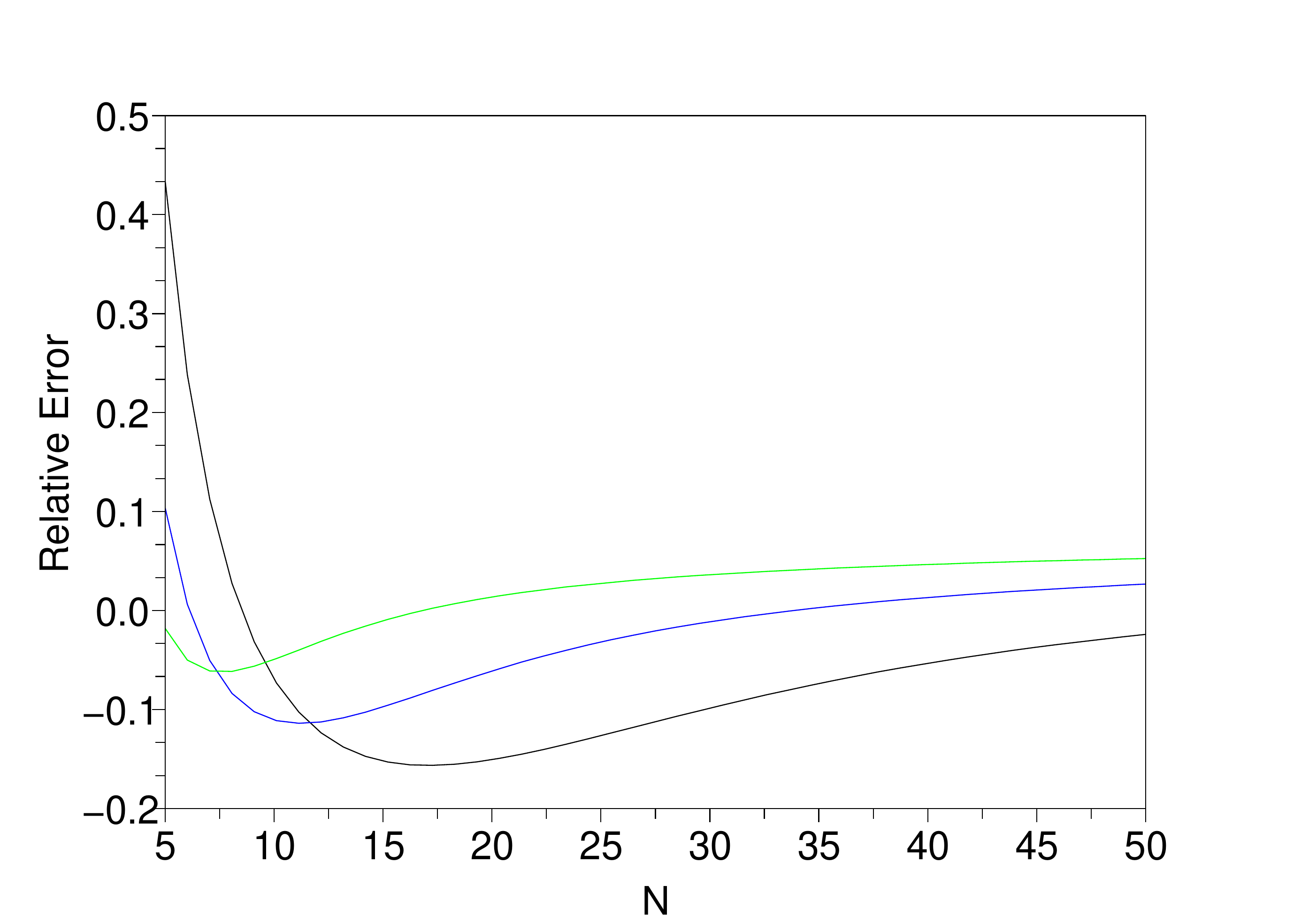}

\caption{Relative error in time to extinction, $(T_{approx}-T)/T$ versus carrying
capacity (population) $N$, for R=1.5 (black line), R=2.0 (blue line)
and R=3.5 (green line) . Small $R$ values are the least accurate
(uppermost) lines at the $N=5$ intercepts. }

\end{figure}

In summary, the asymptotic results from the steepest-descent method
are accurate to within a few percent at large carrying capacity, but
are not reliable below the critical carrying capacity, where exact
results or direct simulations can be used.

These asymptotic results dramatically show that while large carrying
capacity increases extinction times exponentially, it is not the only
factor involved. There is also an exponential dependence on the reproductive
ratio. For a given growth rate and carrying capacity, populations
are exponentially shorter-lived at small reproductive ratio, as $R\rightarrow1$
.

In summary, with carrying capacity above the critical value, there
is a quasi-stable population at $x^{+}\approx g/c$. A population
of this size can last an exponentially long time, although it is not
globally stable. There is a saddle-point, or local maximum in the
potential at a minimum critical population of $x=x^{-}$. If fluctuations
occur below this minimum critical population of $x^{-}\approx b/2g$,
the population is too low to be sustainable. This means that extinction
becomes likely over a short time-scale. Extinction is exponentially
much more rapid with small carrying capacity $N_{c}$, and also with
a small reproductive ratio $R$, which leads to large demographic
fluctuations.

\section{Stochastic discrete logistic model}

So far we have only included demographic noise, due to the intrinsic
random behavior of discrete jump events. In addition to this, there
are external fluctuations, due to variations in environmental parameters
like temperature that affect food supply or other factors relevant
to survival. In our approach, these are represented as random, ie
stochastic, time-dependent rate constants $k(t)$. In general, knowledge
of these rate constants and their statistical fluctuations would allow
more accurate predictions of average extinction times. We know that
the rates have a finite correlation time; no changes are instantaneous.
Despite this, it is useful to treat the limit of a short correlation
time relative to the average growth rate $g$. It is this limit that
we consider here due to its analytic simplicity. More general results
are possible that include the effects of finite correlation times,
but will not be included here.

Effects of this type are sometimes treated by including fluctuating
terms at the rate equation level, without demographic noise due to
discrete events. This has the drawback that the logistic model with
time-varying rates cannot lead to extinction. It simply has no absorbing
state. 

This problem may be circumvented approximately by assuming that a
small population - say $x=1$ - is equivalent to extinction. However,
as we have shown in the previous section, demographic noise itself
plays a role in causing fluctuations leading to extinctions, even
in a static environment. Further, the critical population leading
to a high likelihood of demographic extinction is not necessarily
at $x=1$, but instead is at $x^{-}\approx b/2g$. This depends critically
on the ratio of birth to growth rates. Assuming that $x=1$ is equivalent
to extinction can therefore lead to a serious over-estimation of extinction
times if death and birth rates nearly cancel. 

In the following section we develop a unified theory that includes
both environmental and demographic fluctuation effects in calculating
the extinction rate.

\subsection{Environmental noise }

We now include a specific model for the environmental noise. Environmental
parameters are the time-dependant rates in the logistic equation,
which are positive, and physically have finite bandwidth fluctuations.
These can be modeled by writing the environmental rate parameters
as $k_{i}(t)=k_{i}^{0}\exp(\phi_{i}(t))$. Here $\phi_{i}$ describes
the fluctuations, and is modeled using a stochastic differential equation
of form: \begin{equation}
d\phi=-\gamma_{i}\left[\phi dt+\sigma_{i}dW_{i}^{env}\right]\,\,,\end{equation}
where $dW_{i}$ is an external environmental noise-source such that
$\left\langle dW_{i}dW_{j}\right\rangle =\delta_{ij}dt$, and $\gamma_{i},\sigma_{i}$
describe the rate of change and the relative noise variance of the
$i-th$ rate parameter. The resulting master equation with time-dependent
rates can be termed the stochastic discrete logistic equation, or
SDLE.

For analysis, the time-varying rates can simply be inserted directly
into the demographic Stratonovich equation, (\ref{eq:demographicStratEq}).
This can be numerically integrated with any model of environmental
stochasticity.

If the environmental fluctuation time-scales $\gamma_{i}^{-1}$ are
much smaller than the demographic time-scales $g^{-1}$, adiabatic
elimination will result in an approximate broad-band stochastic noise
with variance $(k_{i}^{0}\sigma_{i})^{2}\delta(t)$. Provided this
variance is relatively small compared to the rate, the fluctuations
can then be linearized to give additive environmental noise in the
rates. 

While one can solve the stochastic equations numerically without these
simplifications, the result is more readily treated analytically,
and will be treated in detail in the remainder of this section. As
a simple example of this, we will consider a fluctuating death rate
with environmental noise variance in the broad-band limit given by:
\begin{equation}
\left\langle \delta a(t)\delta a(t')\right\rangle \simeq2\sigma\delta(t-t')\,\,.\end{equation}

We emphasize that broad-band external noise of this type must be included
in the Stratonovich form of the demographic stochastic equation. Unlike
the Ito form, the Stratonovich equation has a well-defined broad-band
limit with an external noise source. The relative effect of the environmental
noise depends on the competition term $c$, so we will define $r=\sigma/c$
as the \emph{relative environmental noise.}

\subsection{Unified Stratonovich equations}

Adding a fluctuating death rate to the demographic equation in Stratonovich
form, Eq (\ref{eq:demographicStratEq}), we obtain: \begin{equation}
dx^{S}=\left(\left[g+c-\delta a(t)\right]x-b/2-cx^{2}\right)dt+\sqrt{2\left(bx-cx^{2}\right)}dW\,\,,\label{eq:environmentalStratEq1}\end{equation}
Because the demographic and environmental fluctuations are assumed
to be independent, their variances can be added. As both competition
and environmental noise are now broad-band noise sources proportional
to $x$, we introduce $\tilde{c}$ to describe the combined noise
variance, where \begin{equation}
\tilde{c}=c-\sigma=c(1-r)\,\,.\end{equation}
It would not be correct to make this change in (\ref{eq:demographicItoEq}),
the demographic Ito stochastic equation; this does not correspond
to the broad-band limit of a finite bandwidth stochastic process.

Environmental noise has the opposite effect to demographic noise from
density-dependent competition. It increases the variance to super-Poissonian
levels, while intra-specific competition reduces the variance. The
resulting unified Stratonovich equation with demographic \emph{and}
environmental noise is:

\begin{equation}
dx^{S}=\left(\left[g+c\right]x-b/2-cx^{2}\right)dt+\sqrt{2\left(bx-\tilde{c}x^{2}\right)}dW\,\,,\label{eq:UnifiedStratEq}\end{equation}

As previously, we assume absorbing boundary conditions at $x=0$ .
This means that a negative value of $x$ will not occur. For positive
$\tilde{c}$, there is a reflecting boundary at $x=b/\tilde{c}$.
Since $x<b/\tilde{c}$, the equation always remains real. If $\tilde{c}$
is negative, the upper boundary is at infinity. 

A typical simulation of these equations is given in Figure 6. The
dimensionless parameter values are $r=0.75$, $N_{c}=10$, $R=2$.
These simulations use an RK4 (Runge-Kutta) algorithm, with step-size
$\Delta t=0.08$. There are $1000$ trajectories in the ensemble averages.
The mean extinction time, using the method of Eq (\ref{eq:StochasticExtinctionPoisson}),
was calculated to be $T=29.18\pm0.9$ , with time-scales chosen so
that $g=1$. 

\begin{figure}
\includegraphics[angle=270,width=1\textwidth]{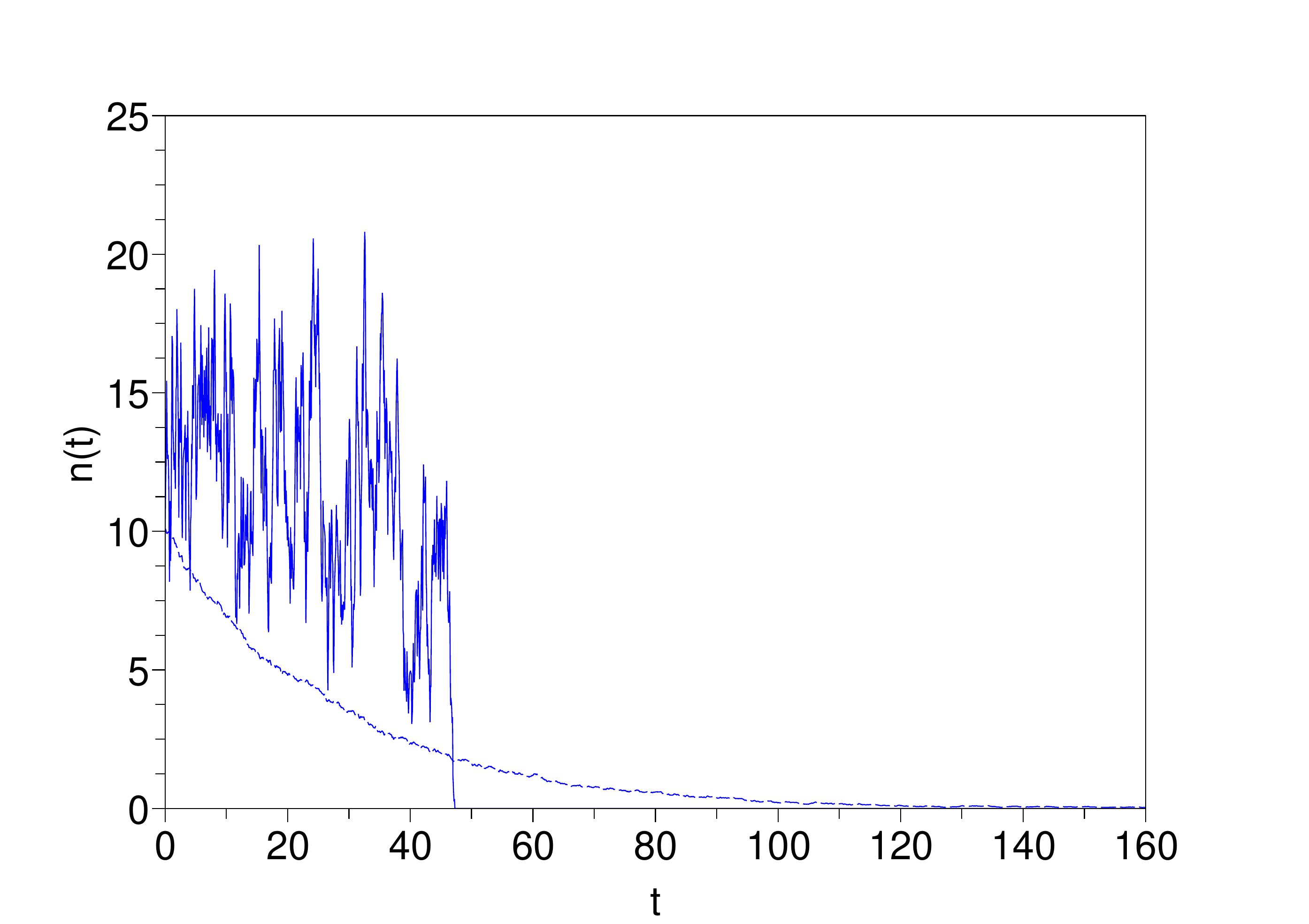}

\caption{Graph of Poisson population versus time: mean values given by the
dotted line, sample simulation given by the solid line, including
environmental noise with $r=0.75$, $N=10$, $R=2$. }

\end{figure}

\subsection{Extinction times}

We now wish to analyse the combined effects of demographic and environmental
noise on extinction times. This is obtained by transforming the unified
Stratonovich equation (\ref{eq:UnifiedStratEq}) back into an Ito
stochastic equation using Eq (\ref{eq:ItoStratTransformation}), and
hence to a Fokker-Planck equation of form: \begin{eqnarray}
\frac{\partial}{\partial t}f\left(x\right) & = & \frac{\partial}{\partial x}\left[\frac{cx-g-\sigma}{\left(b-\tilde{c}x\right)}+\frac{\partial}{\partial x}\right]x\left(b-\tilde{c}x\right)f\left(x\right)\,\,.\nonumber \\
 & = & \frac{\partial}{\partial x}\left[-\tilde{A}(x)+\frac{\partial}{\partial x}\tilde{D}(x)\right]f\left(x\right)\,\,.\end{eqnarray}

We note from Eq (\ref{eq:ItoStratTransformation}), that the transformation
introduces an additional term in the drift that originates from the
environmental noise term in the Stratonovich equation. The quasi-stationary
(steady-state) solution is then:\begin{equation}
\tilde{\psi}\left(x\right)=\frac{1}{x\left(b-\tilde{c}x\right)}\exp\left[\int_{0}^{x}\left(\frac{1}{1-r}-\frac{\tilde{a}}{b-\tilde{c}x}\right)dx\right]\,.\end{equation}

Here we have introduced a new parameter $\tilde{a}\equiv b/(1-r)-cr-g$.
This equation can be integrated to give: 

\begin{equation}
\tilde{\psi}\left(x\right)=\frac{e^{x/(1-r)}}{x}\left(b-\tilde{c}x\right)^{\tilde{a}/\tilde{c}-1}\,.\label{eq:quasi-steady-state}\end{equation}

As with pure demographic fluctuations, an exact extinction time result
is obtainable using: \begin{eqnarray}
\tilde{T}\left(x_{0}\right) & = & 2\int_{0}^{x_{0}}\frac{dx}{\tilde{D}(x)\tilde{\psi}(x)}\int_{x}^{x_{m}}\tilde{\psi}(z)\left(1-e^{-z}\right)dz\,.\label{eq:ExtinctionTimeEnv}\end{eqnarray}

We note that for $\sigma>c$, one formally has $x_{m}=\infty$ in
this treatment. This is an artifact of the use of a Gaussian noise
source, in which there is a small probability of an anomalous, negative
death rate. However, as the large $N$ tails of the distribution have
little or no effect on the extinction time, we expect this to be relatively
unimportant. It should be remarked that the detailed behaviour of
the tails of distribution of death rate for large positive values
are much more significant, especially if there are appreciable departures
from Gaussian statistics. Applying this result to the parameters given
in Figure 6, we obtain $T=28.2109$, within one standard deviation
of the stochastic simulation results.

This excellent agreement between analytic results and simulations
demonstrates that one does not have to use this exact theory. In fact
it may be better not to in some cases. If the environmental noise
has known statistical properties different to those assumed here,
it is preferable to use the simulations described above, with appropriate
noise sources having the true environmental statistics.

\subsection{Steepest Descent Approximation}

The quasi-steady-state distribution is still double-peaked but with
appreciably smaller peak potentials when $\sigma\ge c$. Approximate
results are obtained as before by calculating the extinction time
as an escape probability, from the deterministic or locally stable
value through to extinction at $x=0$. We change to a new variable
$\tilde{y}$, with constant diffusion rate. This variable change is
now defined so that:\begin{equation}
\frac{dx}{d\tilde{y}}=\tilde{\Delta}=\sqrt{x\left(b-\tilde{c}x\right)}\,\,.\end{equation}
 The Fokker-Planck equation for the corresponding probability distribution
$g(\tilde{y})=\Delta f(x)$ is:\begin{equation}
\frac{\partial}{\partial t}g(\tilde{y})=\left[\frac{\partial}{\partial\tilde{y}}V_{\sigma}'(\tilde{y})+\frac{\partial}{\partial\tilde{y}}\right]g\left(\tilde{y}\right)\,,\end{equation}
where $V_{\sigma}'(\tilde{y})=dV_{\sigma}/d\tilde{y}$ , and $V_{\sigma}(\tilde{y})$
is the potential for the new distribution. This is obtained from the
quasi-steady-state distribution - since \begin{eqnarray}
V_{\sigma}(\tilde{y}) & = & -\ln\left(\tilde{\Delta}f_{\infty}\right)\nonumber \\
 & = & -x/(1-r)+\frac{1}{2}\ln\left[x\left(b-\tilde{c}x\right)^{1-2\tilde{a}/\tilde{c}}\right]\,\,.\end{eqnarray}
Alternatively, we can make a variable change to the Stratonovich form
of the drift, which gives the derivative in an equivalent but simpler
form as:\begin{equation}
V_{\sigma}'(\tilde{y})=\frac{1}{\tilde{\Delta}}\left[cx^{2}-(g+c)x+\frac{b}{2}\right]\,\,.\end{equation}

The potential has turning points at $V_{\sigma}'(\tilde{y}^{\pm})=0$
, that are exactly the same as in the pure demographic case. They
correspond to the points where the drift vanishes in the Stratonovich
form of the equation, which is not changed by environmental noise,
so that:\begin{equation}
x^{\pm}=\frac{g+c\pm\gamma}{2c}\,\,,\end{equation}
where $\gamma\equiv\sqrt{(g-c)^{2}-2bc}$ as in the purely demographic
case. 

To obtain the extinction time, it is also necessary to calculate the
curvature of the potential, which is, surprisingly, unchanged from
the purely demographic case:

\begin{eqnarray}
V_{\sigma}"(\tilde{y}^{\pm}) & = & \tilde{\Delta}\frac{d}{dx}V_{\sigma}'(\tilde{y}^{\pm})\nonumber \\
 & = & \pm\gamma\,\,.\end{eqnarray}
The average extinction time is now given by:\begin{equation}
T_{e}=\frac{2\pi}{\gamma}\exp\left[V_{\sigma}(x^{-})-V_{\sigma}(x^{+})\right]\,\,.\end{equation}

This result includes both environmental and demographic contributions
to the extinction time in a single expression. Just as in the case
of pure demographic noise, we can simplify this expression by taking
the large N limit, while fixing the relative level of external noise.
To simplify the resulting expressions, it is convenient to define
an effective reproductive ratio, defined as $\tilde{R}=b/(b+(r-1)g)$.
For a stationary environment, $\tilde{R}=R$. This, of course, is
not the actual reproductive ratio, but rather a scaled parameter that
reflects the relative size of fluctuations in birth events.

This gives rise to the following simple result, which clearly reduces
to the purely demographic result at $r=1$:

\begin{eqnarray}
T & = & T_{0}\exp\left\{ \frac{N_{c}}{1-r}\left[1-\ln\left(\tilde{R}\right)/\left(\tilde{R}-1\right)\right]\right\} +O\left(1/N_{c}\right)\nonumber \\
 & = & T_{0}\exp\left\{ N_{eff}\right\} \,.\end{eqnarray}

As before, we can define a fundamental time-scale of $T_{0}$, and
an \emph{effective} carrying capacity $N_{eff}$ which determines
the exponent: \begin{eqnarray}
T_{0} & = & \frac{2\pi\tilde{R}^{1/(1-r)}}{g}\sqrt{\frac{1-r}{2Ne(\tilde{R}-1)}}\nonumber \\
N_{eff} & = & \frac{N_{c}}{1-r}\left[1-\ln\left(\tilde{R}\right)/\left(\tilde{R}-1\right)\right].\label{eq:envasymptotform}\end{eqnarray}
 Just as in the previous section, one can consider what happens at
large and small $\tilde{R}$, which gives a leading order asymptotic
result of:\begin{eqnarray}
\lim_{\tilde{R}-1\ll1}\ln T=N_{eff} & \rightarrow & \frac{N_{c}}{2(1-r)}\left(\tilde{R}-1\right)+O\left(\ln N_{c}\right)\nonumber \\
\lim_{\tilde{R}\gg1}\ln T=N_{eff} & \rightarrow & N_{c}/(1-r)+O\left(\ln N_{c}\right)\,.\end{eqnarray}

Graph showing typical results are given  in Figures 7 and 8 for relative
environmental fluctuations of $r=1.5$ and $r=3$ respectively. We
see that extinction rates for carrying capacity of $N=20$ are now
three orders of magnitude more rapid than with demographic effects
alone; this relative discrepancy is even larger at higher carrying
capacity. 

\begin{figure}
\includegraphics[angle=270,width=1\textwidth]{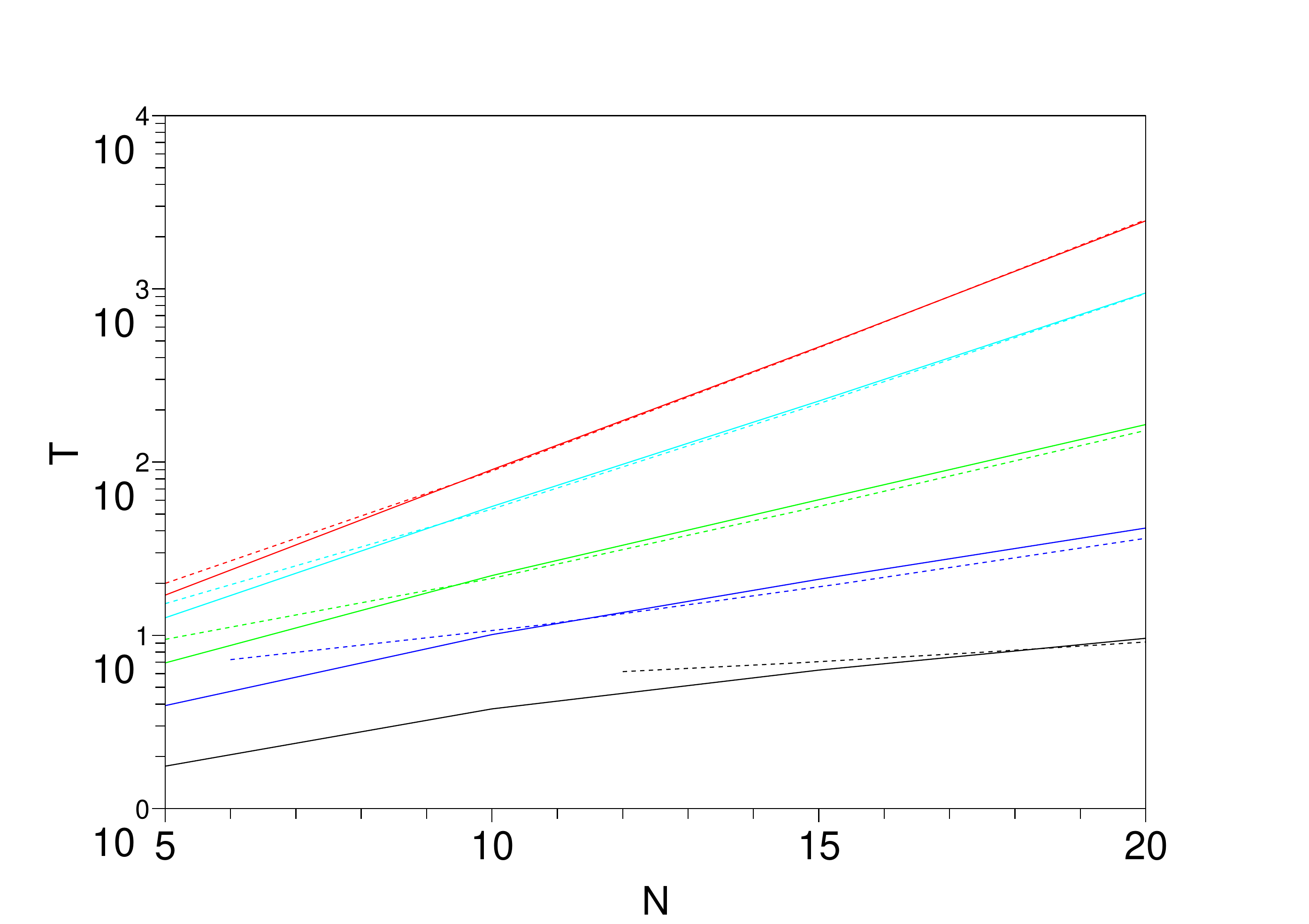}

\caption{Graph of scaled time to extinction versus carrying capacity (population)
$N$, for the same range of $R$ values as in the table, but including
environmental noise with $r=1.5$. Starting from the lower lines,
$R=1.2,1.5,2,3.5,6$. Solid lines are the exact results. Asymptotic
results using steepest descent from Eq (\ref{eq:envasymptotform})
are shown with dashed lines. }

\end{figure}

A clear feature of the results is that environmental fluctuations
have the greatest relative effect on species with a large reproductive
ratio, which otherwise would have an extremely long extinction time
in a static environment. These long extinction times are reduced by
many orders of magnitude, even with relatively small environmental
noise. This is because environmental noise effects can easily be much
larger than the low level of population fluctuations purely due to
demographic causes with large $R$ values. Since environmental fluctuations
are practically unavoidable, we see that the exceptionally long lifetimes
found with large reproductive ratios are probably not achievable in
real world environments. %
\begin{figure}
\includegraphics[angle=270,width=1\textwidth]{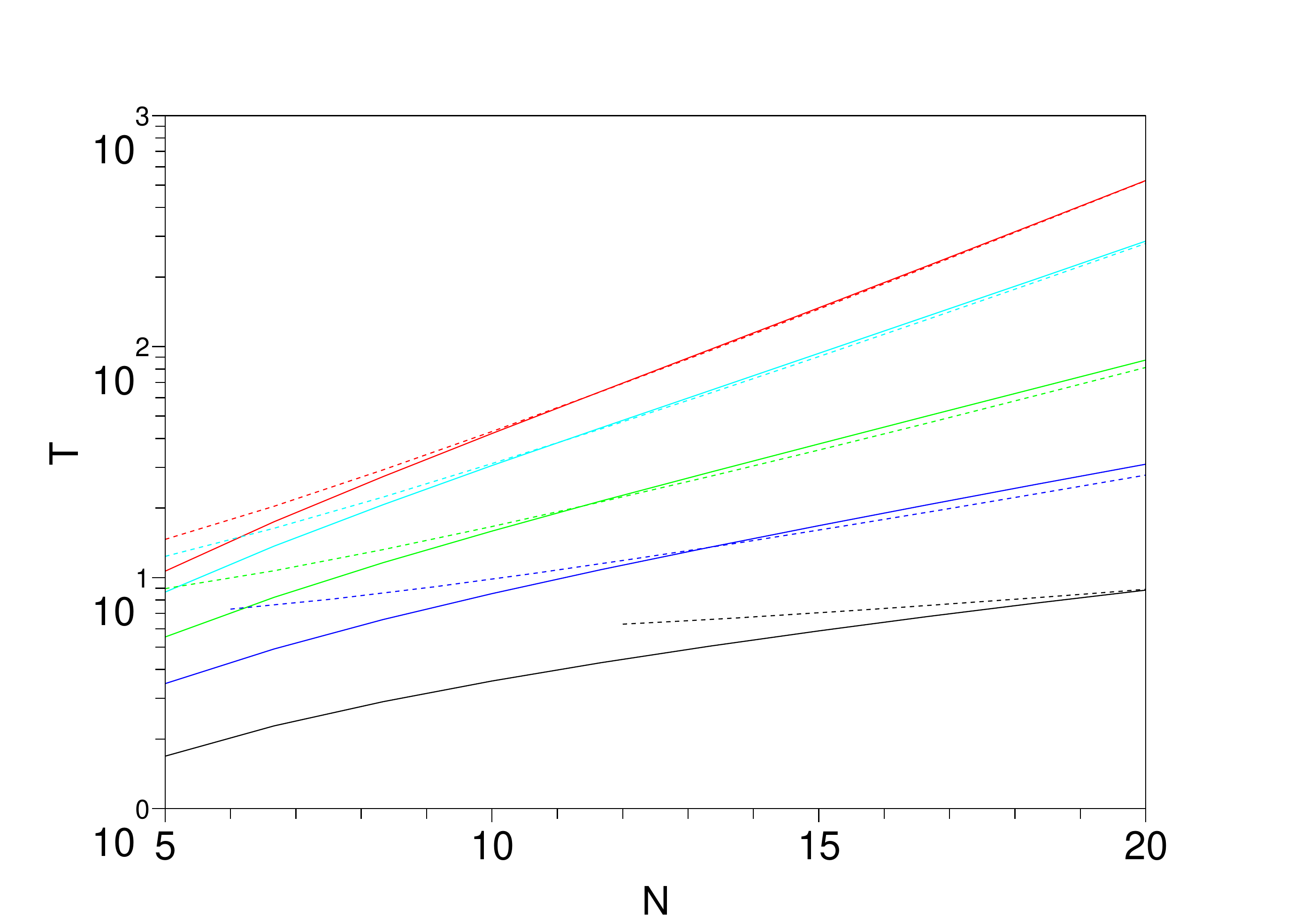}

\caption{Graph of scaled time to extinction versus carrying capacity (population)
$N$, for the same range of $R$ values as in the table, but including
environmental noise with $r=3$. Starting from the lower lines, $R=1.2,1.5,2,3.5,6$.
Solid lines are the exact results. Approximate results using steepest
descent from Eq (\ref{eq:envasymptotform}) are shown with dashed
lines. Parameters are the same as in Fig 3. }

\end{figure}

The effects of varying environmental noise at fixed carrying capacity
$N$ are shown in Figure 9. This shows the strong effects of environmental
noise at large reproductive ratio. It also demonstrates that the much
smaller extinction times found with low reproductive ratios are not
as sensitive to these external effects. This is because small $R$
values mean high death and birth rates, so that the fundamental demographic
noise is high. In this situation, extinction is always rapid. Hence
the faster extinctions due to environmental noise are not so dramatic,
although still very significant when the noise is increased further.

\begin{figure}
\includegraphics[angle=270,width=1\textwidth]{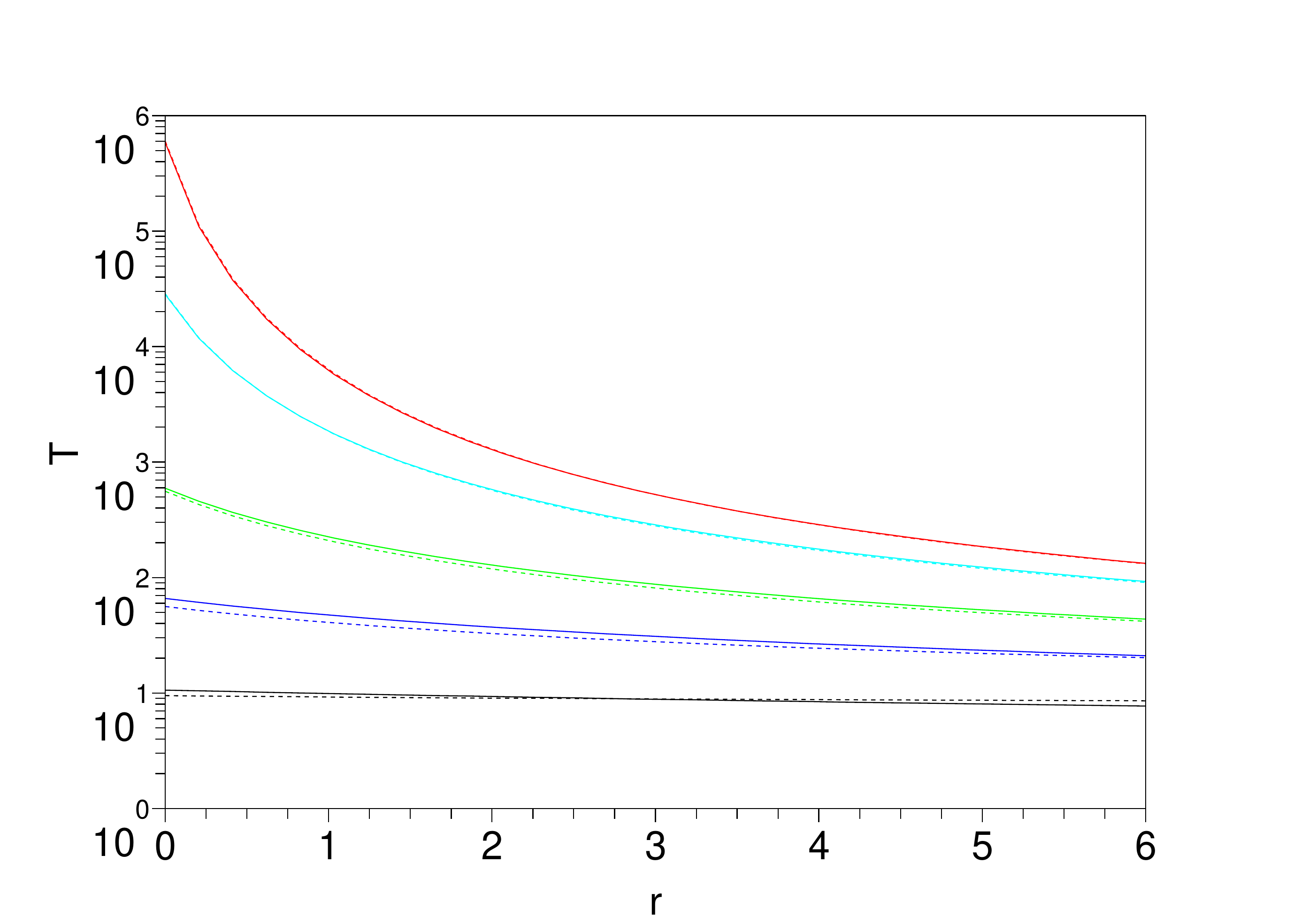}

\caption{Graph of scaled time to extinction versus relative environmental noise
$r$ for the same range of $R$ values as in the table, at a carrying
capacity of $N=20$. Starting from the lower lines, $R=1.2,1.5,2,3.5,6$.
Solid lines are the exact results. Approximate results using steepest
descent from Eq (\ref{eq:envasymptotform}) are shown with dashed
lines. Parameters are the same as in Fig 3.}

\end{figure}

\section*{Discussion}

We have described a simple class of models that can be used to describe
an isolated self-regulating population, including both demographic
and environmental fluctuations. We believe the stochastic discrete
logistic model represents an appropriate basis for a new synthesis
of population genetics, mathematical epidemiology and theoretical
ecology. We have used three equivalent techniques for analysing population
dynamics and extinction times: direct master equation simulations,
stochastic equations and exactly soluble Fokker-Planck equations. 

We emphasize that in the Poisson representation used here, all three
techniques are exact and give identical results. This is to be contrasted
with previous work using truncated forms of the Fokker-Planck equation,
in which there can be exponentially large errors introduced by the
diffusion approximation.

The Fokker-Planck method has the useful feature that it allows precise
yet analytically tractable calculations of the asymptotic extinction
times for large carrying capacity. At large $N_{c,}$ we find a general
dependence of $\log T\propto N_{c}$, but with a very different constants
depending on the reproductive ratio $R$ and the relative environmental
noise $r$. A single analytic expression agrees to within a few percent
of the exact extinction times over a wide range of parameter values,
if the carrying capacity is not too low. Provided these parameter
values can be estimated, this provides a useful basis for risk analysis
of survival probabilities in small, isolated populations.

Our results show that extinction times have an exponential dependence
on carrying capacity above a critical carrying capacity that depends
on the reproductive ratio. Surprisingly, the effect of small reproductive
ratio is just as serious as small carrying capacity, in that both
cause exponential reductions in extinction time to the point of almost
total non-viability in the short term. 

Environmental fluctuations have a similar effect, although these are
much more pronounced when extinction rates are low, as occurs with
large $R$ values. These results show clearly that carrying capacity
or observed population sizes in isolated ecosystems are not by themselves
a reliable guide to long-term viability. Instead, the total picture
of reproductive ratio and fluctuations in growth rates due to environmental
causes needs to be included as well in any risk assessment.

There are many ways in which this work can be extended. The most obvious
is to extend this analysis to include genetics so that extinction
would represent the loss of an allele from the population. By introducing
mutation, a revised neutral theory of evolution could be developed
that accommodated demographic fluctuations directly. Similarly natural
selection could be modeled either through differential growth rates,
differential death rates or inter-genotype competition.

A more challenging direction of enquiry will be to develop a framework
for inference under the spatial logistic model analogous to that which
Kingman's coalescent provides for analyzing the idealized Wright-Fisher
and Moran population models.

\section*{Acknowledgements}

We would like to thank Peter Wills for useful discussions. PDD was
supported in this research by an Australian Research Council Discovery
Grant.

\section*{Figure Captions}

\paragraph*{Figure 1 }

Direct simulation results using Gillespie algorithm with parameters
$b=2$, $a=1$, $c=0.2$ . This corresponds to $N_{c}=5$, $R=2$,
$g=1$. The initial condition of $N_{0}=8$ was sampled from a Poissonian
with mean of $x_{0}=5$. The solid line is a single stochastic realization,
showing integer jump behaviour. The dotted line is an average of $10000$
realizations, showing the exponential decline in average population
leading to extinction.

\paragraph*{Figure 2}

Poisson simulation results using Stratonovich equations with parameters
$b=2$, $a=1$, $c=0.2$ or $R=2$, $N_{c}=5$ as in Figure 1. The
initial condition was a Poissonian with mean of $x_{0}=5$. The solid
line is a single stochastic realization, showing continuous stochastic
behaviour. The dotted line is an average of $10000$ realizations,
showing the exponential decline in average population leading to extinction,
just as in the Gillespie result. Integration step-size utilized was
$dt=0.04$, RK4 Runge-Kutta algorithm. Average extinction time $T=10.04\pm0.27$
using Eq(\ref{eq:StochasticExtinctionPoisson})

\paragraph*{Figure 3}

Dimensionless time to extinction versus carrying capacity (population)
$N$, for the same range of $R$ values as in the table. Starting
from the lower lines, $R=1.2,1.5,2,3.5,6$. Solid lines are exact
from Eq(\ref{eq:ExtinctionTime}), dashed line are the steepest descent
result from Eq (\ref{eq:DemSteepest}), dotted lines are asymptotic
(large $N$) expressions from Eq (\ref{eq:Asymptotdemographic}).

\paragraph*{Figure 4}

Dimensionless time to extinction versus reproductive ratio $R$, for
$N=10,\,20$ (lower and upper curves respectively). Solid line is
the exact result, dashed line is the asymptotic expression (\ref{eq:Asymptotdemographic})

\paragraph*{Figure 5}

Relative error in time to extinction, $(T_{approx}-T)/T$ versus carrying
capacity (population) $N$, for R=1.5 (black line), R=2.0 (blue line)
and R=3.5 (green line) . Small $R$ values are the least accurate
(uppermost) lines at the $N=5$ intercepts.

\paragraph*{Figure 6}

Graph of Poisson population versus time: mean values given by the
dotted line, sample simulation given by the solid line, including
environmental noise with $r=0.75$, $N=10$, $R=2$.

\paragraph*{Figure 7}

Graph of scaled time to extinction versus carrying capacity (population)
$N$, for the same range of $R$ values as in the table, but including
environmental noise with $r=1.5$. Starting from the lower lines,
$R=1.2,1.5,2,3.5,6$. Solid lines are the exact results. Asymptotic
results using steepest descent from Eq (\ref{eq:envasymptotform})
are shown with dashed lines.

\paragraph*{Figure 8}

Graph of scaled time to extinction versus carrying capacity (population)
$N$, for the same range of $R$ values as in the table, but including
environmental noise with $r=3$. Starting from the lower lines, $R=1.2,1.5,2,3.5,6$.
Solid lines are the exact results. Approximate results using steepest
descent from Eq (\ref{eq:envasymptotform}) are shown with dashed
lines. Parameters are the same as in Fig 3.

\paragraph*{Figure 9}

Graph of scaled time to extinction versus relative environmental noise
$r$ for the same range of $R$ values as in the table, at a carrying
capacity of $N=20$. Starting from the lower lines, $R=1.2,1.5,2,3.5,6$.
Solid lines are the exact results. Approximate results using steepest
descent from Eq (\ref{eq:envasymptotform}) are shown with dashed
lines. Parameters are the same as in Fig 3.

\bibliographystyle{elsart-harv}
\bibliography{Library}

\end{document}